\documentclass[pra,onecolumn,floatfix,a4paper,superscriptaddress]{revtex4}
\usepackage{bm,color,graphicx,amsmath,txfonts}

\usepackage[colorlinks, citecolor=blue,linkcolor=blue]{hyperref}

\begin{document}
\title{Monogamy of Gaussian quantum steering and entanglement in a hybrid qubit-cavity optomagnonic system with coherent feedback loop}

\author{Hamza Harraf}
\affiliation{LPHE-Modeling and Simulation, Faculty of Sciences, Mohammed V University in Rabat, Rabat, Morocco}	
\author{Mohamed Amazioug} 
\thanks{m.amazioug@uiz.ac.ma}
\affiliation{LPTHE-Department of Physics, Faculty of sciences, Ibnou Zohr University, Agadir, Morocco}
\author{Amjad Sohail}
\affiliation{Department of Physics Government College University Allama Iqbal Road, Faisalabad 38000, Pakistan}
\affiliation{Instituto de Física Gleb Wataghin Universidade Estadual de Campinas Campinas, SP 13083-859, Brazil}	
\author{Rachid Ahl Laamara}
\affiliation{LPHE-Modeling and Simulation, Faculty of Sciences, Mohammed V University in Rabat, Rabat, Morocco}	
\affiliation{Centre of Physics and Mathematics, CPM, Faculty of Sciences, Mohammed V University in Rabat, Rabat, Morocco}	
	
\begin{abstract}
	
The monogamy of quantum correlations is a fundamental principle in quantum information processing, limiting how quantum correlations can be shared among multiple subsystems. Here we propose a theoretical scheme to investigate the monogamy of quantum steering and genuine tripartite entanglement in a hybrid qubit-cavity optomagnonic system with a coherent feedback loop. Using logarithmic negativity and Gaussian quantum steering, we quantify entanglement and steerability, respectively. We verify the CKW-type monogamy inequalities which leads to steering monogamous through adjustments of the reflective parameter among three tripartite modes versus temperature. Our results show that a coherent feedback loop can enhance entanglement and quantum steering under thermal effects.

\end{abstract}
\date{\today}
\maketitle
	
\section{Introduction} 

Quantum entanglement, a fundamental resource for quantum information, has been extensively studied across diverse physical systems \cite{intro1}. Its applications range from testing foundational quantum mechanics \cite{intro2,intro3} to enabling quantum information processing \cite{intro4} and building quantum networks \cite{intro5}. A particularly intriguing aspect is \textit{Einstein-Podolsky-Rosen (EPR) steering} \cite{intro6}, a strict subset of entanglement that has attracted significant attention due to its applications in quantum key distribution \cite{intro7}. First proposed by Schr\"odinger in response to the EPR paradox \cite{intro8}, steering exhibits a unique asymmetric property: one observer can influence another's quantum state through local measurements, distinguishing it from both Bell nonlocality and standard entanglement \cite{intro9}.

EPR steering has been demonstrated in various systems, including atom-mechanical \cite{intro10}, optomechanical \cite{intro11,intro12}, and waveguide-coupled architectures \cite{intro13}. Remarkably, one-way steering can be achieved in unbroken $\mathcal{PT}$-symmetric systems \cite{intro14}, while tripartite steering shows enhanced robustness in one-sided device-independent scenarios compared to two-sided configurations \cite{intro15}. Recent advances include the experimental realization of high-dimensional steering using orbital-angular-momentum-encoded photons \cite{intro16} and the study of $N$-qubit entanglement with genuine $N$-partite steering in loophole-free settings \cite{intro17}. 

In optomechanical systems, radiation pressure enables the generation of entanglement between optical cavities and macroscopic mirrors \cite{intro18}, with demonstrated capabilities for entanglement transfer \cite{intro19}. These features make such systems particularly promising for quantum information applications.

In recent years, cavity magnonics has attracted growing interest and achieved significant advances \cite{intro20}. Magnons the quanta of collective spin excitations in magnetic materials can coherently interact with diverse quantum systems, offering a promising platform for hybrid quantum technologies \cite{intro21}. These include magnon dark mode memories \cite{intro22}, single-magnon detection \cite{intro23}, quantum thermodynamics \cite{thermo24}, magnon-magnon entanglement \cite{PLA24}, magnon squeezing \cite{Fan24}, magnon blockade \cite{AdP24}, and strongly coupled magnon-photon systems. In particular, the magnetic dipole interaction in yttrium iron garnet (YIG) spheres has enabled both theoretical and experimental demonstrations of strong magnon-photon coupling \cite{intro24,Li2018} and magnon-photon entanglement \cite{intro25}. Further developments have shown efficient magnon-superconducting qubit interactions \cite{intro26,intro30}, while magnetostrictive interactions in YIG spheres facilitate magnon-phonon coupling. This mechanism enables applications such as parametric amplification \cite{intro27} and magnon-mediated photon-phonon conversion \cite{intro28}. Recent work on cavity optomagnonic systems has exploited magneto-optical whispering gallery modes (WGMs) \cite{intro29}, yet studies of quantum entanglement and EPR steering in such WGM systems remain scarce—an important gap this work addresses.

We investigate a hybrid qubit-cavity optomagnonic system that consists of a YIG sphere and a superconducting qubit embedded within a single-mode microwave cavity. We employ a coherent feedback technique to enhance the system's coupling effects, a method previously explored in \cite{intro30}. The system's dynamics are governed by three primary interactions: a magneto-optical coupling between the cavity mode and the YIG sphere mediated by whispering gallery modes (WGMs); an electric dipole interaction between the cavity mode and the superconducting qubit, which produces a beamsplitter-type coupling; and a parametric interaction between the cavity and magnon modes. This specific configuration allows the cavity mode to mediate an indirect coupling between the qubit and magnon modes. The quantum correlations, including entanglement and steerability, are generated and enhanced through the coherent feedback loop. We identify and categorize three distinct types of steerability---one-way, two-way, and no-way steering---and also characterize the monogamy of quantum steering and genuine tripartite entanglement within the system.

This paper is organized as follows. In Section~\ref{sec1}, we derive the Hamiltonian of the qubit-cavity optomagnonic system and the corresponding quantum Langevin equations (QLEs) with a coherent feedback loop. Section \ref{sec2} we evaluate the explicit formula of the covariance matrix (CV). Section~\ref{sec3} establishes the entanglement and Gaussian quantum steering between the indirectly coupled qubit and magnon modes. In Section~\ref{sec4}, we present numerical results analyzing how entanglement and Gaussian quantum steering depend on the system parameters. Finally, Section~\ref{sec5} provides our concluding remarks.

\section{Model} \label{sec1}

\begin{figure}[ht!]
\centering
\begin{tabular}{ccc}
\includegraphics[width=0.51\textwidth]{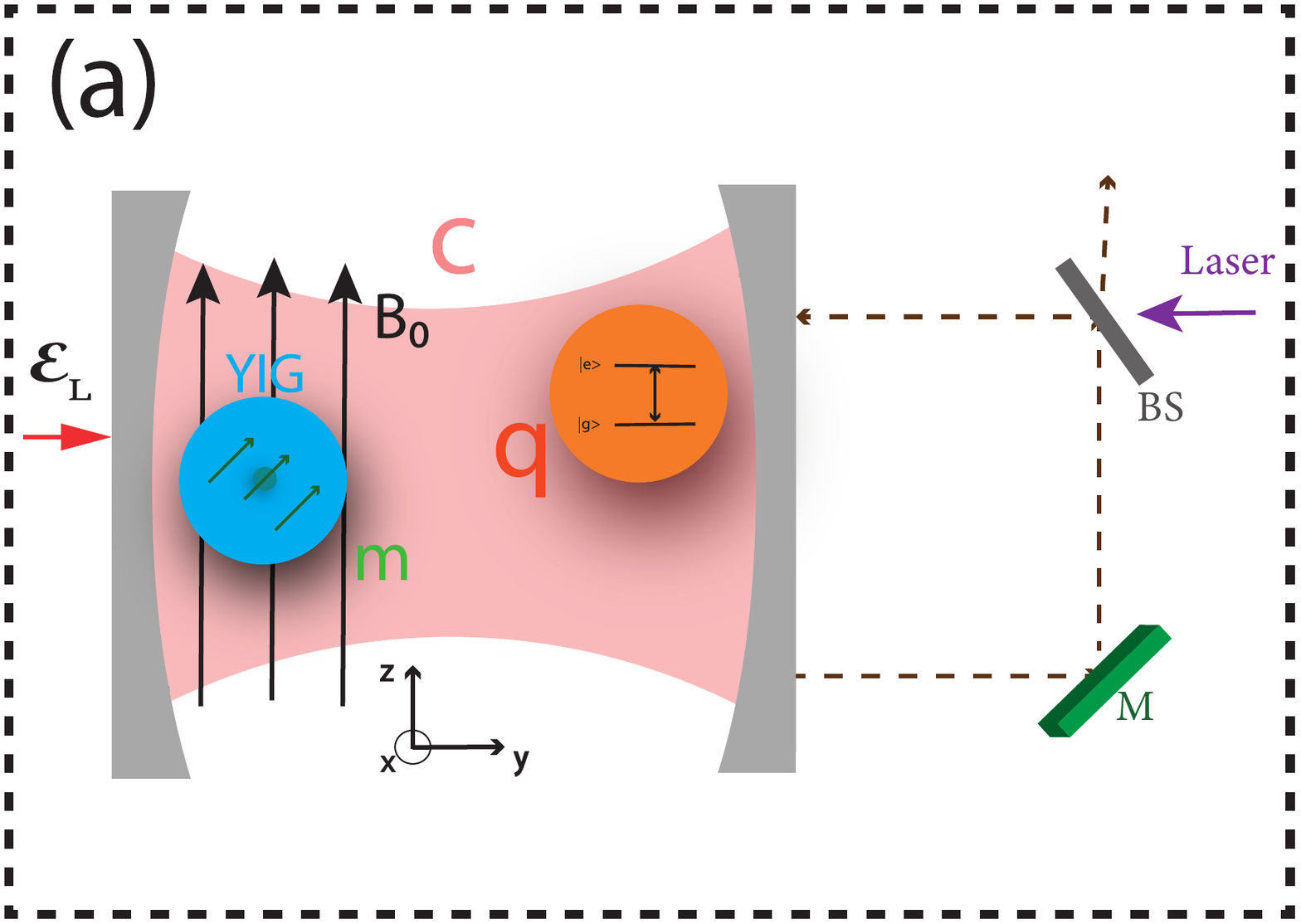} 
\hspace{0.30cm}
\includegraphics[width=0.51\textwidth]{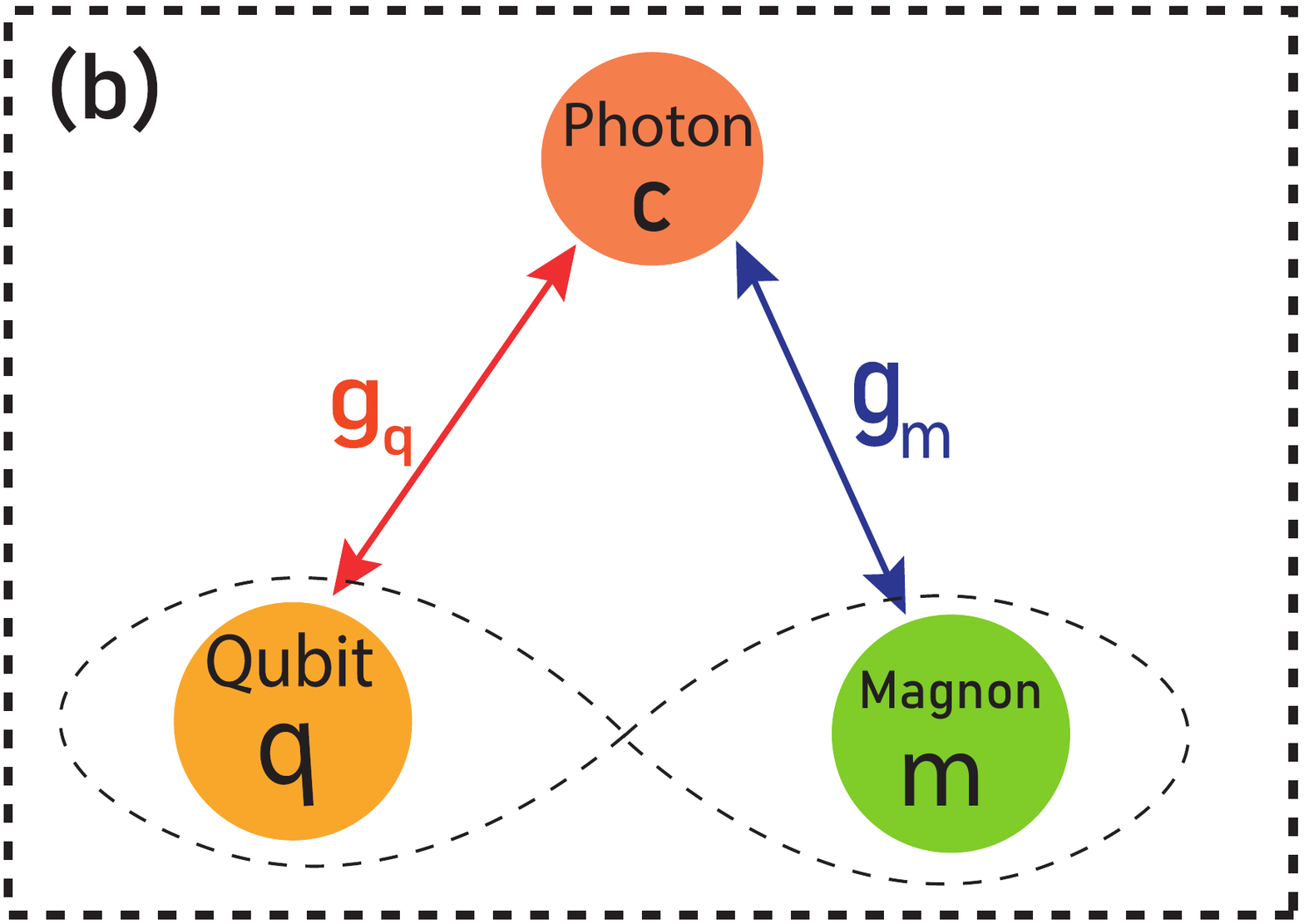}
\end{tabular}
\caption{(a) A schematic of the hybrid cavity-qubit optomagnonic system using coherent feedback loop. The ferromagnetic YIG sphere, which contains the collective motion of spins representing the magnons, is placed inside the cavity's optical whispering gallery mode (WGM) ($c$). An external magnetic field $B_0$ is applied along the z-axis. The cavity is driven by an input electromagnetic field with amplitude $\mathcal{A}$ through an asymmetric beam splitter (BS) with transmission and reflection coefficients $u$ and $\epsilon$, respectively. (b) The interaction among the subsystems. The cavity is coupled to the superconducting qubit ($q$) with a coupling strength $g_q$ and to the magnon mode ($m$) with a coupling strength $g_m$.}
\label{fig1}
\end{figure}

We investigate a cavity-qubit optomechanic system composed of a three-dimensional single-mode optical cavity, a superconducting qubit, and a spherical yttrium iron garnet (YIG) ferrimagnetic crystal. As illustrated in Fig. \ref{fig1}(a), the system includes a coherent feedback protocol \cite{Harwood2021}. A superconducting qubit couples to the optical cavity mode with a Rabi frequency $g_{q}$ \cite{model1,Ning2021}, and the ferrimagnetic YIG couples to the optical cavity mode with a coupling strength $g_{m}$, as shown in Fig. \ref{fig1}(b). The ferromagnetic YIG material has a high spin density and a low damping rate, which allows for both an optical resonant mode through magneto-optical WGMs \cite{model2} and a homogeneous Kittel magnon mode \cite{model3,model4}. When the magnetic component of the cavity field is perpendicular to the bias magnetic field, various magnetostatic modes are activated in the YIG sphere. We focus on the uniform Kittel magnon mode and ignore other phonon modes due to the WGM coupling in our optomagnonic system \cite{model5, model6}. The Kittel mode is magnetized by a magnetic field $B_0$ along the z-direction and displays consistent spin precessions across the entire YIG sphere volume. It also has the strongest magnetic coupling with the microwave cavity, a reasonable assumption since the microwave wavelength is much larger than the sphere's size. The frequency of the Kittel mode $\Omega_{m}=\Gamma_{0}B_{0}$, is determined by the gyromagnetic ratio ($\Gamma_{0}=2\pi\times28\times10^{9}$ Hz/T) and the applied magnetic field $B_0$. Applying the Holstein-Primakoff approximation \cite{model7}, we represent the magnon mode with boson creation ($m^{\dagger}$) and annihilation ($m$) operators. The superconducting qubit is likewise represented by bosonic operators ($q$) and ($q^{\dagger}$). To increase the optomagnonic coupling strength, a laser field with amplitude $\mathcal{E}_{L}$ and frequency $\Omega_{L}$ drives the cavity mode \cite{model8, model9}.

The total Hamiltonian of the system, $\mathcal{H}_T$, can be expressed in the bosonic mode form as
\begin{align}
\label{eq1}
\mathcal{H}_T/\hbar=&\Omega_{c}c^{\dagger}c+\Omega_{q}q^{\dagger}q+\Omega_{m}m^{\dagger}m\\ \nonumber
              +&g_{m}c^{\dagger}c(m+m^{\dagger})+g_{q}(cq^{\dagger}+c^{\dagger}q)\\ \nonumber
              +&(\mathcal{E}_Lc^{\dagger}\mathrm{e}^{-i\Omega_Lt}+\mathcal{E}^{*}_Lc \mathrm{e}^{i\Omega_Lt})+u\mathcal{A}(c^{\dagger}\mathrm{e}^{i\beta}+c\mathrm{e}^{-i\beta}), \nonumber
\end{align}
where, $c^{\dagger}$ and $c$ symbolize the creation and annihilation operators for the cavity mode, with a frequency of $\Omega_c$. The frequency of the qubit mode is denoted by $\Omega_q$, and the coupling strength between the cavity and the qubit is $g_q$. Experimental demonstrations of the coupling between the optical and magnon modes have been realized in optomagnonic systems \cite{model10,model11}, where optical photons and magnons interact with a coupling strength
\begin{equation}\label{eq1_2}
g_{m}=V\frac{c}{n_{\rm{r}}}\sqrt{\frac{2}{n_{\rm{s}}V_{\rm{sp}}}},
\end{equation}
where the Verdet constant of YIG is given by $V= 3.77\times 10^{2}$ rad.m$^{-1}$, the refractive index of the material is $n_{\rm{r}}=2.19$, and the spin density is $n_{\rm{s}}=2.1\times 10^{28}/\text{m}^3$. The volume of the YIG sphere is $V_{\rm{sp}}=4\frac{\pi}{3} r^3$, with a radius of $r=10^{-4}\text{m}$, and $c=3\times 10^{8}\text{m/s}$ is the speed of light in a vacuum. The final term in equation (\ref{eq1}) describes the optical field transmitted through the beam splitter \cite{model16,FBAdP}. The parameters are: $\beta$, the phase of the electromagnetic field; $\mathcal{A}$, the amplitude of the coherent laser source; and $u$ and $\epsilon$, the real amplitude transmission and reflection coefficients of the beam splitter. These coefficients are positive and satisfy the relation $u^{2}+\epsilon^{2}=1$. If there is no coherent feedback of the output field into the cavity, then $\epsilon=0$ and $u=1$. The Hamiltonian in equation (\ref{eq1}) has an explicit time dependence. We can transform it into a time-independent form using the unitary transformation $\mathcal{H}=\mathcal{U}^{\dagger}\mathcal{H}_T\mathcal{U}+i\hbar\frac{d\mathcal{U}}{dt}\mathcal{U}^{\dagger}$, where $\mathcal{U}(t)=\exp[i\Omega_L(c^{\dagger}c+q^{\dagger}q)t]$.
\begin{align}\label{eq2}
\mathcal{H}/\hbar&=\Delta_{c0} c^{\dagger}c+\Delta_{q}q^{\dagger}q+\Omega_{m} m^{\dagger}m\\ \nonumber
&+g_{m}c^{\dagger}c(m+m^{\dagger}) +g_{q}(cq^{\dagger}+c^{\dagger}q)\\ \nonumber 
&+(\mathcal{E}_Lc^{\dagger}+\mathcal{E}^{*}_Lc)+u\mathcal{A}(c^{\dagger}\mathrm{e}^{i\beta}+c\mathrm{e}^{-i\beta}).\nonumber
\end{align}
Here, $\Delta_{c0}=\Omega_{c}-\Omega_L$ and $\Delta_q=\Omega_q-\Omega_L$ are the frequency detunings for the microwave and qubit modes, respectively, with respect to the driving field. To linearize the Hamiltonian, we represent the mode operators as the sum of a large classical amplitude and a small fluctuation (noise) operator. That is, $\Psi=\bar{\Psi}+\delta\Psi$, where $\Psi$ can be the operator for the cavity ($c$), magnon ($m$), or qubit ($q$) mode
\begin{align}
\mathcal{H}_n/\hbar&=\Delta_{c0}\delta c\delta c^{\dagger}+\Delta_{q}\delta q\delta q^{\dagger}+\Omega_{m}\delta m^{\dagger}\delta m\\ \nonumber
&+g_m|\bar{c}|(\delta c+\delta c^{\dagger})(\delta m+\delta m^{\dagger})+g_{m}( \bar{m}+ \bar{m}^{\dagger})\delta c\delta c^{\dagger}+ g_q(\delta c\delta q^{\dagger}+\delta q\delta c^{\dagger})\\ \nonumber
&+u\mathcal{A}(\delta c^{\dagger}\mathrm{e}^{i\beta}+\delta c\mathrm{e}^{-i\beta}).
\end{align}
The Hamiltonian can be writes as
\begin{align}\label{eq3}
\mathcal{H}_n/\hbar&=\Delta_{c}\delta c^{\dagger}\delta {c}+\Delta_{q} \delta q^{\dagger}\delta q+\Omega_{m}\delta m^{\dagger}\delta m \\ \nonumber
&+\tilde{g}_{m}(\delta c+\delta c^{\dagger})(\delta m+\delta m^{\dagger})\\ \nonumber
&+g_{q}(\delta c\delta q^{\dagger}+\delta c^{\dagger}\delta q)+u\mathcal{A}(\delta c^{\dagger}\mathrm{e}^{i\beta}+\delta c\mathrm{e}^{-i\beta}),\nonumber
\end{align}
with $\Delta_c = \Delta_{c0}+g_m(\bar{m}+\bar{m}^{\dagger})$ represents the effective detuning of the cavity, and $\tilde{g}_m = g_m|\bar{c}|$ is the effective optomagnonic coupling constant \cite{model12}. The average amplitude of the optical field $|\bar{c}|$, is given by $|\bar{c}|=\sqrt{\frac{2\mathsf{P}_p}{k_c\hbar\Omega_P}}$, where $\mathsf{P}_p$ is the optical power \cite{model13}. Using equation (\ref{eq3}), we can derive the quantum Langevin equations (QLEs) that describe the dynamics of the hybrid system
\begin{align}\label{(eq4)}
\frac{dc}{dt}&=-(k_{\text{fb}}+i\Delta_{\text{fb}})c-i\tilde{g}_{m}(m+m^{\dagger})-
ig_{q}q-iu\mathcal{A}\mathrm{e}^{i\beta}+\sqrt{2k_{c}}c_{\text{fb}}^{\text{in}},\\ \nonumber
\frac{dq}{dt}&=-(\Gamma+i\Delta_{q})q-ig_{q} c+\sqrt{2\Gamma}q^{\text{in}},\\ \nonumber
\frac{dm}{dt}&=-(k_{m}+i\Omega_{m})m-i\tilde{g}_{m}(c+c^{\dagger})+\sqrt{2k_{m}}m^{\text{in}}.
\end{align}
The damping rates for the cavity, magnon, and qubit modes are denoted by $k_{\text{fb}}=k_{c}(1-2\epsilon\cos(\theta))$, $k_{m}$, and $\Gamma$, respectively. The effective detuning of the cavity mode is $\Delta_{\text{fb}}=\Delta_{c}-2k_{c}\epsilon\sin(\theta)$. The phase shift from the output field's reflection on the mirrors is $\theta$, and the input quantum noises, $q^{\text{in}}$ and $m^{\text{in}}$, are characterized by zero correlations and average values \cite{model14}. The effective input noise operator, which exists due to coherent feedback, is given by $C_{\text{fb}}^{\text{in}}=\epsilon \mathrm{e}^{i\theta}c^{\text{out}}+u c^{\text{in}}$. The noise operator for the microwave mode, $c^{\text{in}}$ is the only one with non-zero correlations in this system \textcolor{red}{\cite{amazioug2023enhancement,Li2017}}. Additionally, the standard input-output relation $c^{\text{out}}=\sqrt{2k_{c}}c-uc^{\text{in}}$ establishes a relationship between the output field $c^{\text{out}}$ and the cavity field $c$. This leads to the effective input noise operator $C^{\text{in}}_{\text{fb}}=\epsilon\sqrt{2k_{c}}e^{i\theta}c+c_{\text{fb}}^{\text{in}}$. Furthermore, for the cavities, the non-zero coherent feedback correlations of the input noise operators $c^{\text{in}}_{\text{fb}}$ and $c^{\text{in}\dagger}_{\text{fb}}$ are given by $c_{\text{fb}}^{\text{in}}=u(1-\epsilon e^{i\theta})c^{\text{in}}$, and their non-zero correlation functions are
\begin{equation}\label{eq5}
 \begin{aligned} 
\langle c^{\text{in}}_{\text{fb}}(t)c^{\text{in}\dagger}_{\text{fb}}(t')\rangle &=\{u^2(1-\epsilon\mathrm{e}^{i\theta})(1-\epsilon\mathrm{e}^{-i\theta})\}[N_{c}(\Omega_{c})+1]\delta(t-t'), \\ 
\langle c^{\text{in}}_{\text{fb}}(t)c^{\text{in}\dagger}_{\text{fb}}(t')\rangle &=\{u^2(1-\epsilon\mathrm{e}^{i\theta})(1-\epsilon\mathrm{e}^{-i\theta})\}[N_{c}(\Omega_{c})]\delta(t-t').
\end{aligned}
\end{equation}
\begin{equation}\label{eq6}
\langle \nu^{\text{in}}(t)\nu^{\text{in}\dagger}(t')\rangle=[N_{\nu}+1]\delta(t-t'),\quad \langle \nu^{\text{in}\dagger}(t)\nu^{\text{in}}(t')\rangle=N_{\nu}\delta(t-t'),\quad  \nu^{\text{in}}=q^{\text{in}},m^{\text{in}}.
\end{equation}
The average number of thermal excitations in the modes is $N_{\eta}=\frac{1}{\left[\exp\left(\frac{\hbar\Omega_{\eta}}{k_{b}T}\right)-1\right]}$ for $\eta=c,m,q$. To produce entanglement, at least one coupling between the modes in a bipartite system must be parametric. The chosen coupling type depends on the detunings. We will now convert the operators into rotating frames
$
\tilde{c}=c\mathrm{e}^{i\Delta_{\text{fb}}t}$, $ \tilde{q}=q\mathrm{e}^{i\Delta_{q}t}$ and $\tilde{m}=m\mathrm{e}^{i\Omega_{m}t},
$
the equations (\ref{(eq4)}) is given by
\begin{align}\label{eq8}
\frac{d\tilde{c}}{dt}&=-k_{\text{fb}}\tilde{c}-i\tilde{g}_{m}\left(\tilde{m}\mathrm{e}^{i(\Delta_{\text{fb}}-\Omega_{m})t}+\tilde{m}^{\dagger}\mathrm{e}^{i(\Delta_{\text{fb}}+\Omega_{m})t}\right)-ig_{q}\tilde{q}\mathrm{e}^{i(\Delta_{\text{fb}}-\Delta_{q})t}-iu\mathcal{A} \mathrm{e}^{i(\Delta_{\text{fb}}t+\beta)}+\sqrt{2k_{c}}c^{\text{in}}_{\text{fb}},\\ \nonumber
\frac{d\tilde{q}}{dt}&=-\Gamma\tilde{q}-ig_{q}\tilde{c}\mathrm{e}^{i(\Delta_{q}-\Delta_{\text{fb}})t}+\sqrt{2\Gamma}q^{\text{in}},\\ \nonumber
\frac{d\tilde{m}}{dt}&=-k_{m}\tilde{m}-i\tilde{g}_{m}\left(\tilde{c}^{\dagger}
\mathrm{e}^{i(\Delta_{\text{fb}}+\Omega_{m})t}+\tilde{c}\mathrm{e}^{-i(\Delta_{\text{fb}}-\Omega_{m})t}\right)+\sqrt{2k_{m}}m^{\text{in}}.
\end{align}
It is clear that the modes' frequencies can be adjusted to either $+\Omega_{m}$ or $-\Omega_{m}$, which are referred to as the blue and red sidebands, respectively. We observe that by appropriately tuning the detuning, we can change the nature of the coupling between the modes. Consequently, by selecting the pairing with the blue sideband ($\Delta_{\text{fb}}\approx \Delta_{q}\approx -\Omega_{m}$) and assuming $\Delta_{\text{fb}}t\ll\beta$, while ignoring rapidly oscillating terms with $\pm 2\Omega_{m}$, we can obtain the QLEs
\begin{align}
\label{eq9}
\frac{d\tilde{c}}{dt}&=-k_{\text{fb}}\tilde{c}-i\tilde{g}_{m}\tilde{m}^{\dagger}-ig_{q}\tilde{q}-iu\mathcal{A}\mathrm{e}^{i\beta}+\sqrt{2k_{c}}c^{\text{in}}_{\text{fb}},\\ \nonumber
\frac{d\tilde{q}}{dt}&=-\Gamma\tilde{q}-ig_{q}\tilde{c}+\sqrt{2\Gamma}q^{\text{in}},\\ \nonumber
\frac{d\tilde{m}}{dt}&=-k_{m}\tilde{m}-i\tilde{g}_{m} \tilde{c}^{\dagger}+\sqrt{2k_{m}}m^{\text{in}}.
\end{align}

\section{Covariance matrix}  \label{sec2}
We define the quadrature operators for the cavity, qubit, and magnon modes as 
$\tilde{X}_{c}=\frac{\tilde{c}+\tilde{c}^{\dagger}}{\sqrt{2}}$, $\tilde{Y}_{c}=\frac{\tilde{c}-\tilde{c}^{\dagger}}{i\sqrt{2}}$; $\tilde{X}_{q}=\frac{\tilde{q}+\tilde{q}^{\dagger}}{\sqrt{2}}$, $\tilde{Y}_{q}=\frac{\tilde{q}-\tilde{q}^{\dagger}}{i\sqrt{2}}$; and $\tilde{X}_{m}=\frac{\tilde{m}+\tilde{m}^{\dagger}}{\sqrt{2}}$, $\tilde{Y}_{m}=\frac{\tilde{m}-\tilde{m}^{\dagger}}{i\sqrt{2}}$, respectively. By using equation (\ref{eq9}), a set of quantum Langevin equations (QLEs) is derived for the quadrature operators
\begin{align}
\frac{d\tilde{X}_{c}}{dt}=&-k_{\text{fb}}\tilde{X}_{c}+g_{q}\tilde{Y}_{q}-\tilde{g}_{m}\tilde{Y}_{m}+\sqrt{2k_{c}}X_{c}^{\text{in}},\\ \nonumber
\frac{d\tilde{Y}_{c}}{dt}=&-k_{\text{fb}}\tilde{Y}_{c}-g_{q}\tilde{X}_{q}-\tilde{g}_{m}\tilde{X}_{m}+\sqrt{2k_{c}}Y^{\text{in}}_{c},\\ \nonumber
\frac{d\tilde{X}_{q}}{dt}=&-\Gamma \tilde{X}_{q}+g_{q}\tilde{Y}_{q}+\sqrt{2\Gamma}q^{\text{in}},\\ \nonumber
\frac{d\tilde{Y}_{q}}{dt}=&-\Gamma \tilde{Y}_{q}-g_{q}\tilde{X}_{q}+\sqrt{2\Gamma}Y^{in}_{q},\\ \nonumber
\frac{d\tilde{X}_{m}}{dt}=&-k_{m} \tilde{X}_{m}-\tilde{g}_{m}\tilde{Y}_c+\sqrt{2k_{m}}X_{m}^{\text{in}},\\ \nonumber
\frac{d\tilde{Y}_{m}}{dt}=&-k_{m} \tilde{Y}_{m}-\tilde{g}_{m}\tilde{X}_{c}+\sqrt{2k_{m}}Y^{\text{in}}_{m}.\\ \nonumber
\end{align}
The QLEs can be arranged into a matrix form as follows
\begin{equation}\label{eq10}
\dot{\lambda}(t)=\mathcal{Q}\lambda(t)+\mathsf{n}(t),
\end{equation}
where $\mathsf{n}(t)^T=[\sqrt{2k_{c}}X^{\text{in}}_{c},\sqrt{2k_{c}}Y^{\text{in}}_{c},\sqrt{2\Gamma}X^{\text{in}}_{q},\sqrt{2\Gamma}Y^{\text{in}}_{q},\sqrt{2k_{m}}X^{\text{in}}_{m},\sqrt{2k_{m}}Y^{\text{in}}_{m}]$ and $\lambda(t)^T=[ \tilde{X}_{c}, \tilde{Y}_{c}, \tilde{X}_{q}, \tilde{Y}_{q}, \tilde{X}_{m}, \tilde{Y}_{m}]$, In addition, the drift matrix $\mathcal{Q}$ is expressed by
\begin{equation}
\mathcal{Q}=
\begin{pmatrix}
-k_{\text{fb}}&0&0&g_{q}&0&-\tilde{g}_{m}\\
0&-k_{\text{fb}}&-g_{q}&0&-\tilde{g}_{m}&0\\
0&g_{q}&-\Gamma&0&0&0\\
-g_{q}&0&0&-\Gamma&0&0\\
0&-\tilde{g}_{m}&0&0&-k_{m}&0\\
-\tilde{g}_{m}&0&0&0&0&-k_{m}\\
\end{pmatrix}.
\end{equation}
Owing to the Gaussian nature of the quantum noise, the system's steady-state dynamics (described by the QLEs in equation (\ref{eq10})) can be characterized by a $6\times 6$ covariance matrix (CM), $\mathcal{V}$. The steady-state CM is determined by solving the Lyapunov equation \cite{model15}
\begin{equation}
\mathcal{Q}\mathcal{V}+\mathcal{V}\mathcal{Q}^T=-\mathcal{D},
\end{equation}
with $\mathcal{D}=\text{diag}[k_{c}u^2(1-\epsilon)^2(2N_{c}+1),k_{c}u^2(1-\epsilon)^2(2N_{c}+1),\Gamma(2N_{q}+1),\Gamma(2N_{q}+1),k_{m}(2N_{m}+1),k_{m}(2N_{m}+1)]$.
The steady-state covariance matrix of our system can therefore be expressed as follows
\begin{equation}
	\mathcal{V}=
	\begin{pmatrix}
	v_{11}& 0 & 0& v_{14} &0 &v_{16} \\
	0  &v_{11}&-v_{14}&0 &v_{16} &0\\
	0&-v_{14}&v_{33}&0&v_{35}&0\\
	v_{14}&0&0&v_{33}&0&-v_{35}\\
	0&v_{16}&v_{35}&0&v_{66}&0\\
	v_{16}&0&0&-v_{35}&0&v_{66}\\
	\end{pmatrix},
\end{equation}
\begin{align}
v_{11} &= \left[-u^2 (1-\epsilon)^2 k_{c} \left(\Gamma \tilde{g}_{m}^4 + \left(g_{q}^2 + \Gamma (\Gamma + k_{\text{fb}})\right) k_{m} \left(g_{q}^2 + (\Gamma + k_{m}) (k_{\text{fb}} + k_{m})\right) \right.\right. \nonumber \\
& \quad \left.\left. - \tilde{g}_{m}^2 \left(\Gamma^3 + g_{q}^2 (\Gamma + k_{m}) + \Gamma k_{m} (\Gamma + k_{m}) + \Gamma k_{\text{fb}} (\Gamma + 2 k_{m})\right)\right) (1 + 2 N_{c})\right. \nonumber \\
& \left.\quad + \tilde{g}_{m}^2 k_{m} \left(-\Gamma \tilde{g}_{m}^2 + \Gamma^2 (\Gamma + k_{\text{fb}}) + \left(g_{q}^2 + \Gamma (\Gamma + k_{\text{fb}})\right) k_{m}\right) (1 + 2 N_{m})\right. \nonumber \\
& \quad \left.+ \Gamma g_{q}^2 \left(-\Gamma \tilde{g}_{m}^2 + k_{m} \left(g_{q}^2 + (\Gamma + k_{m}) (k_{\text{fb}} + k_{m})\right)\right) (1 + 2 N_{q}) \right] \nonumber \\
& \quad /\left[ 2 \left(\Gamma \tilde{g}_{m}^2 - \left(g_{q}^2 + \Gamma k_{\text{fb}}\right) k_{m}\right) \left(g_{q}^2 (\Gamma + k_{\text{fb}}) + (k_{\text{fb}} + k_{m}) \left(-\tilde{g}_{m}^2 + (\Gamma + k_{\text{fb}}) (\Gamma + k_{m})\right)\right)\right],
\end{align}
\begin{flushleft}
 \begin{align}
v_{14}=&\left[g_{q} \left(\Gamma u^2 (1-\epsilon)^2 k_{c} \left(-\Gamma \tilde{g}_{m}^2 + k_{m} \left(g_{q}^2 + \left(\Gamma + k_{m}\right) \left(k_{\text{fb}} + k_{m}\right)\right)\right) \left(1 + 2 N_{c}\right) \right.\right.  \nonumber \\
&\left. + \Gamma \tilde{g}_{m}^2 k_{m} \left(\Gamma + k_{\text{fb}} + k_{m}\right) \left(1 + 2 N_{m}\right) \right. \nonumber \\
&\left. \left.+ \Gamma \left(\tilde{g}_{m}^2 \left(\Gamma + k_{m}\right) \left(k_{\text{fb}} + k_{m}\right) - k_{\text{fb}} k_{m} \left(g_{q}^2 + \left(\Gamma + k_{m}\right) \left(k_{\text{fb}} + k_{m}\right)\right)\right) \left(1 + 2 N_{q}\right)\right)\right] \nonumber\\
&\quad /\left[ 2 \left(\Gamma \tilde{g}_{m}^2 - \left(g_{q}^2 + \Gamma k_{\text{fb}}\right) k_{m}\right) \left(g_{q}^2 (\Gamma + k_{\text{fb}}) + (k_{\text{fb}} + k_{m}) \left(-\tilde{g}_{m}^2 + (\Gamma + k_{\text{fb}}) (\Gamma + k_{m})\right)\right)\right],
\end{align}
   \end{flushleft}
   \begin{flushleft}
   \begin{align}
v_{16}=&\left[\tilde{g}_{m} g_{q} \left(u^2 (1-\epsilon)^2 k_{c} \left(-\Gamma \tilde{g}_{m}^2 + k_{m} \left(g_{q}^2 + \Gamma \left(\Gamma + 2 k_{\text{fb}} + k_{m}\right)\right)\right) \left(1 + 2 N_{c}\right) \right.\right. \nonumber\\
&\left. + \left(g_{q}^2 \left(\Gamma + k_{\text{fb}}\right) + \Gamma \left(\tilde{g}_{m}^2 + k_{\text{fb}} \left(\Gamma + k_{\text{fb}}\right)\right)\right) k_{m} \left(1 + 2 N_{m}\right) \right. \nonumber \\
&\left. + \Gamma \left(g_{q}^2 k_{m} + \tilde{g}_{m}^2 \left(k_{\text{fb}} + k_{m}\right) - k_{\text{fb}} k_{m} \left(k_{\text{fb}} + k_{m}\right)\right) \left(1 + 2 N_{q}\right)\right]\nonumber \\
&\quad/\left[ 2 \left(\Gamma \tilde{g}_{m}^2 - \left(g_{q}^2 + \Gamma k_{\text{fb}}\right) k_{m}\right) \left(g_{q}^2 (\Gamma + k_{\text{fb}}) + (k_{\text{fb}} + k_{m}) \left(-\tilde{g}_{m}^2 + (\Gamma + k_{\text{fb}}) (\Gamma + k_{m})\right)\right)\right],
\end{align}
   \end{flushleft}
     \quad
   \begin{flushleft}
 \begin{align}
v_{33} &=\left[ -u^2 (1-\epsilon )^2 g_{q}^2 k_{c} \left(-\Gamma  \tilde{g}_{m}^2 + k_{m} \left(g_{q}^2 + \left(\Gamma + k_{m}\right) \left(k_{\text{fb}} + k_{m}\right)\right)\right) \left(1 + 2 N_{c}\right)\right. \nonumber \\
& \quad + \tilde{g}_{m}^2 g_{q}^2 k_{m} \left(\Gamma + k_{\text{fb}} + k_{\mathsf{m}}\right) \left(1 + 2 N_{m}\right) \nonumber \\
& \quad + \Gamma \left(g_{q}^4 k_{m} + \left(k_{\text{fb}} + k_{m}\right) \left(-\tilde{g}_{m}^2 + k_{\text{fb}} k_{m}\right) \left(-\tilde{g}_{m}^2 + \left(\Gamma + k_{\text{fb}}\right) \left(\Gamma + k_{m}\right)\right) \right. \nonumber \\
& \quad \left.\left. + g_{q}^2 \left(\tilde{g}_{m}^2 \left(-\Gamma + k_{m}\right) + k_{m} \left(k_{\text{fb}} \left(2 \Gamma + k_{\text{fb}}\right) + \left(\Gamma + k_{\text{fb}}\right) k_{m} + k_{m}^2\right)\right)\right) \left(1 + 2 N_{q}\right)\right] \nonumber \\
& \quad/\left[ 2 \left(\Gamma \tilde{g}_{m}^2 - \left(g_{q}^2 + \Gamma k_{\text{fb}}\right) k_{m}\right) \left(g_{q}^2 (\Gamma + k_{\text{fb}}) + (k_{\text{fb}} + k_{m}) \left(-\tilde{g}_{m}^2 + (\Gamma + k_{\text{fb}}) (\Gamma + k_{m})\right)\right)\right],
\end{align}
   \end{flushleft}
     \quad
   \begin{flushleft}
   \begin{align}
  v_{35}& =\left[\tilde{g}_{m} g_{q} \left(u^2 (1-\epsilon)^2 k_{c} \left(-\Gamma \tilde{g}_{m}^2 + k_{m} \left(g_{q}^2 + \Gamma \left(\Gamma + 2 k_{\text{fb}} + k_{m}\right)\right)\right) \left(1 + 2 N_{c}\right) \right. \right. \nonumber \\
&\left. + \left(g_{q}^2 \left(\Gamma + k_{\text{fb}}\right) + \Gamma \left(\tilde{g}_{m}^2 + k_{\text{fb}} \left(\Gamma + k_{\text{fb}}\right)\right)\right) k_{m} \left(1 + 2 N_{m}\right) \right.  \nonumber \\
&\left. \left.+ \Gamma \left(g_{q}^2 k_{m} + \tilde{g}_{m}^2 \left(k_{\text{fb}} + k_{m}\right) - k_{\text{fb}} k_{m} \left(k_{\text{fb}} + k_{m}\right)\right) \left(1 + 2 N_{q}\right)\right)\right] \nonumber \\
&\quad /\left[ 2 \left(\Gamma \tilde{g}_{m}^2 - \left(g_{q}^2 + \Gamma k_{\text{fb}}\right) k_{m}\right) \left(g_{q}^2 (\Gamma + k_{\text{fb}}) + (k_{\text{fb}} + k_{m}) \left(-\tilde{g}_{m}^2 + (\Gamma + k_{\text{fb}}) (\Gamma + k_{m})\right)\right)\right],
\end{align}
   \end{flushleft}
   \quad
   \begin{flushleft}
   \begin{align}
  v_{66} &= \left[k_{\mathsf{m}} \left(-g_{q}^4 \left(\Gamma + k_{\text{fb}}\right) + \Gamma \left(-\tilde{g}_{m}^2 + \left(\Gamma + k_{\text{fb}}\right) \left(\Gamma + k_{m}\right)\right) \left(\tilde{g}_{m}^2 - k_{\text{fb}} \left(k_{\text{fb}} + k_{m}\right)\right) \right.\right. \nonumber \\
& \quad \left. + g_{q}^2 \left(\tilde{g}_{m}^2 \left(-\Gamma + k_{m}\right) - \left(\Gamma + k_{\text{fb}}\right) \left(2 \Gamma k_{\text{fb}} + \left(\Gamma + k_{\text{fb}}\right) k_{m} + k_{m}^2\right)\right)\right) \left(1 + 2 N_{m}\right) \nonumber \\
& \quad + \tilde{g}_{m}^2 \left(u^2 (1-\epsilon)^2 k_{c} \left(\Gamma \left(\tilde{g}_{m}^2 - \Gamma \left(\Gamma + k_{\text{fb}}\right)\right) - \left(g_{q}^2 + \Gamma \left(\Gamma + k_{\text{fb}}\right)\right) k_{m}\right) \left(1 + 2 N_{c}\right) \right. \nonumber \\
& \quad \left.\left. - \Gamma g_{q}^2 \left(\Gamma + k_{\text{fb}} + k_{m}\right) \left(1 + 2 N_{q}\right)\right)\right] \nonumber \\
& \quad /\left[ 2 \left(\Gamma \tilde{g}_{m}^2 - \left(g_{q}^2 + \Gamma k_{\text{fb}}\right) k_{m}\right) \left(g_{q}^2 (\Gamma + k_{\text{fb}}) + (k_{\text{fb}} + k_{m}) \left(-\tilde{g}_{m}^2 + (\Gamma + k_{\text{fb}}) (\Gamma + k_{m})\right)\right)\right].
\end{align}
\end{flushleft}

\section{Bi- and tripartite quantum correlations}\label{sec3}
\subsection{Quantum entanglement}

We primarily focus on Gaussian bipartite entanglement, which is quantified by the logarithmic negativity ($\mathcal{L}_N$) \cite{model17, model18}. The logarithmic negativity is given by
\begin{equation}
\label{eq23}
\mathcal{L}_N=\max [0,-\ln 2 \vartheta],
\end{equation}
the value of $\vartheta$ is given by $\vartheta=\frac{\sqrt{\sigma-\sqrt{\sigma^2-4\det\mathcal{V}_{XY}}}}{\sqrt{2}}$, where $\sigma=\det\mathcal{X}+\det\mathcal{Y}-2\det\mathcal{Z}$. The matrix $\mathcal{V}_{XY}$ is a $4\times4$ sub-matrix of the covariance matrix $\mathcal{V}$, and its elements depend on the pairwise entanglement of the two modes under consideration. It can be expressed as
\begin{equation}
\mathcal{V}_{XY}=
\begin{pmatrix}
\mathcal{X}&\mathcal{Z}\\
\mathcal{Z}^T&\mathcal{Y}\\
\end{pmatrix},
\end{equation}
where $2\times2$ sub-matrices of $\mathcal{V}_{XY}$ are $\mathcal{X}=\text{diag}(v_{33},v_{33})$, $\mathcal{Y}=\text{diag}(v_{66},v_{66})$, and $\mathcal{Z}=\text{diag}(v_{35},-v_{35})$. We use the residual contangle $\mathbb{R}$ \cite{adesso2007entanglement} as a quantitative measure to examine the tripartite entanglement of the system. The contangle is the continuous-variable (CV) equivalent of tangle, which is used for discrete-variable tripartite entanglement \cite{amazioug2023enhancement, adesso2006continuous}
\begin{equation}
\mathbb{R}_{\text{min}}\equiv \min[\mathbb{R}^{i|jk},\mathbb{R}^{k|ij},\mathbb{R}^{j|ik}],
\end{equation}
with $\mathbb{R}^{i|jk}$ is the residual contangle, while $\mathrm{C}_{\mathsf{u}|\mathsf{v}}$ is the contangle between subsystems $\mathsf{u}$ and $\mathsf{v}$ ($\mathsf{v}$ consists of one or two modes). The contangle $\mathrm{C}_{\mathsf{u}|\mathsf{v}}$ is defined as the squared logarithmic negativity, $\mathrm{C}_{\mathsf{u}|\mathsf{v}}\equiv \mathcal{L}_{\mathsf{u}|\mathsf{v}}^{2}$. A nonzero minimum residual contangle ($\mathbb{R}_{\text{min}}$) indicates the existence of genuine tripartite entanglement. The expression for $\mathbb{R}^{i|jk}$ is similar to the Coffman-Kundu-Wootters monogamy inequality \cite{amazioug2023enhancement,intro31}, which holds for a system of three modes. The residual contangle $\mathbb{R}^{i|jk}$ is given by
\begin{equation}
 \mathbb{R}^{i|jk}=\mathrm{C}_{i|jk}-\mathrm{C}_{i|j}-\mathrm{C}_{i|k}\geq 0,  \quad   (i,j,k=c,m,q),
\end{equation}
we use equation (\ref{eq23}) to calculate the logarithmic negativity $\mathcal{L}_{\mathsf{u}|\mathsf{v}}$ when $\mathsf{v}$ is a single mode (e.g., $\mathcal{L}_{c|q}$, $\mathcal{L}_{c|m}$, and $\mathcal{L}_{q|m}$). For cases where $\mathsf{v}$ represents two modes (e.g., $\mathcal{L}_{c|qm}$, $\mathcal{L}_{q|cm}$, and $\mathcal{L}_{m|cq}$), the $\vartheta$ in equation (\ref{eq23}) is defined by
\begin{equation}
\vartheta =\min \operatorname{eig}\left[i \hat{\Omega}_3 \tilde{\mathcal{V}_{\beta}}\right],
\end{equation}
where 
$\hat{\Omega}_3$ and $\tilde{\mathcal{V}}_{\beta}$ are defined, respectively, as
$
\hat{\Omega}_3=\oplus_{\beta=1}^3 i \sigma_y, \quad \sigma_y=\left(\begin{array}{cc}
	0 & -i \\
	i & 0
\end{array}\right)
$ and $ \tilde{\mathcal{V}}_{\beta}=\mathbb{P}_{ \alpha \mid \beta \gamma} \mathcal{V}_6 \mathbb{P}_{\alpha \mid \beta \gamma}\quad(\alpha \neq \beta \neq \gamma)$, where $\mathcal{V}_6$ again a $6\times 6$ covariance matrix of three correlated modes. 
 Where $\quad \mathbb{P}_{\alpha \mid \beta\gamma}=\sigma_z\oplus 1 \oplus 1, \quad \mathbb{P}_{\beta\mid \alpha \gamma}=$ $1\oplus \sigma_z \oplus 1$, and $\mathbb{P}_{\gamma \mid \alpha \beta}=1\oplus 1 \oplus \sigma_z$, with $\sigma_z=\text{diag}(1,-1)$.

\subsection{Quantum steering}

Another quantum correlation quantifier is Gaussian quantum steering. The steerability of Bob ($X$) by Alice ($Y$) ($X\to Y$) for a ($n_X+n_Y$)-mode Gaussian state can be quantified by \cite{IKogias2015}
\begin{equation} \label{eq:37}
	\mathcal{G}^{X\to Y} (\mathcal{V}_{XY}) = \max\left[0,-\sum_{j:\bar \nu_j^{XY/ Y}<1/2}\ln\left(\bar\nu_j^{XY/ X}\right)\right],
\end{equation}
where the quantity $\bar{\nu}_j^{XY/X}$ ($j=1,...,m_Y$) represents the symplectic eigenvalues of the matrix $\bar{\mathcal{V}}_{XY/X}=\mathcal{Y}-\mathcal{Z}^T\mathcal{X}^{-1}\mathcal{Z}$. This matrix is obtained from the Schur complement of $\mathcal{X}$ in the covariance matrix $\mathcal{V}_{XY}$. The steerability of Alice by Bob, denoted by $[\mathcal{G}^{Y\to X} (\mathcal{V}_{XY})]$, is obtained by swapping the roles of $X$ and $Y$. To assess the asymmetric steerability of the two-mode Gaussian state, we introduce the steering asymmetry, defined as
\begin{equation} \label{eq:37}
\mathcal{G}(XY)=\left|\mathcal{G}^{X\to Y} - \mathcal{G}^{Y\to X}\right|.
\end{equation}
Consequently, we can distinguish between three types of steering: one-way, two-way, and no-way steering. Both symmetrical and asymmetrical versions are encompassed by two-way steering \cite{model21}. The key results are summarized below: First, there is no steering between modes $X$ and $Y$ when both $\mathcal{G}^{X\to Y}=0$ and $\mathcal{G}^{Y\to X}=0$. Second, one-way steering exists from mode $Y$ to mode $X$ when $\mathcal{G}^{Y\to X}>0$ but $\mathcal{G}^{X\to Y}=0$. Third, there is asymmetric two-way steering between modes $X$ and $Y$ when both $\mathcal{G}^{X\to Y}>0$ and $\mathcal{G}^{Y\to X}>0$, but the measures are not equal, i.e., $\mathcal{G}^{X\to Y}\neq \mathcal{G}^{Y\to X}$. Lastly, there is symmetric two-way steering between modes $X$ and $Y$ when $\mathcal{G}^{Y\to X}=\mathcal{G}^{X\to Y}>0$, which means the steering asymmetry $\mathcal{G}(XY)=0$.

Following the criterion from Ref.~\cite{IKogias2015}, we analyze monogamy steering by considering all possible bipartite separations. The Coffman-Kundu-Wootters (CKW)-type monogamy relations \cite{VCoffman2000}, which quantify the distribution of steering among subsystems \cite{YXiang2017}, are given by
\begin{equation} \label{eq:37}
\begin{split}
 \mathcal{G}^{k\to (i, j)}(\mathcal{V}_{ijk}) - \mathcal{G}^{k\to i}(\mathcal{V}_{ijk}) - \mathcal{G}^{k\to j}(\mathcal{V}_{ijk}) \geq 0,\\
 \mathcal{G}^{(i, j)\to k}(\mathcal{V}_{ijk}) - \mathcal{G}^{i\to k}(\mathcal{V}_{ijk}) - \mathcal{G}^{j\to k}(\mathcal{V}_{ijk}) \geq 0.
\end{split}
\end{equation}
For the tripartite continuous-variable Gaussian state, where $i,j,k\in\{X,Y,Z\}$, the Coffman-Kundu-Wootters (CKW)-type monogamy relation is validated for all types of Einstein-Podolsky-Rosen (EPR) steering in the tripartite optomagnomechanical system, as illustrated in Figs.~\ref{fig10} and \ref{fig11}.

\section{Results and discussions} \label{sec4}

In this section, we will discuss the results for steady-state entanglement and Gaussian quantum steering between the different modes of our system, and we will demonstrate the effect of feedback on these properties. Our analysis will focus on how the reservoir temperature and reflictivity parameter influence these interactions. We employ experimentally feasible parameters \cite{intro30}, where $\mathrm{B}_0 = 100 \times 10^{-3} \text{ T}$ is the amplitude of the external bias magnetic field. This field is used to create a magnon with frequency $\Omega_{m}=\Gamma_0 B_0$, where the gyromagnetic ratio is $\Gamma_0=2\pi\times28\times 10^9 \text{ Hz/T}$ and the damping rate is $k_{m}/2\pi= 10^6 \text{ Hz}$. The coupling strength $g_{m}$ is determined using the parameters from Eq.~(\ref{eq1_2}). We selected a YIG Verdet constant of $V=3.77\times10^2$ rad.m$^{-1}$, a refractive index $n_{\rm{r}}=2.19$, a spin density $n_{\rm{s}}=2.1\times10^{28}\text{/m}^3$, and a YIG sphere radius of $r=100\times10^{-6}\text{ m}$. For the microwave mode ($c$), we choose a damping rate of $k_{c}/2\pi=5\times10^6$ Hz and a resonance frequency of $\Omega_{c}/2\pi=8.35\times10^9$ Hz. The driving field has a power of $\mathsf{P}_p=10\times10^{-3}$ W and a wavelength of $\lambda_p=2\pi c/\Omega_p=1550\times10^{-9}$ m. This produces an intra-cavity photon number of $n_p=\frac{2\mathsf{P}_p}{k_{c}\hbar\Omega_p}$ \cite{model22}. The effective optomagnonic coupling constant is $\tilde{g}_m=g_{m}|\sqrt{n_p}|=17.35\times10^6$ Hz \cite{model23}. For the superconducting qubit ($q$), we select a resonance frequency of $\Omega_{q}/2\pi=8.44\times10^9$ Hz and a damping rate of $\Gamma/2\pi=0.2\times10^6$ Hz \citep{model24}. The temperature is set to $T=10\times10^{-3}$ K and the phase is $\theta=\pi$.
\\
\begin{figure}[ht!]
\centering
\begin{tabular}{ccc}
\includegraphics[scale=0.35]{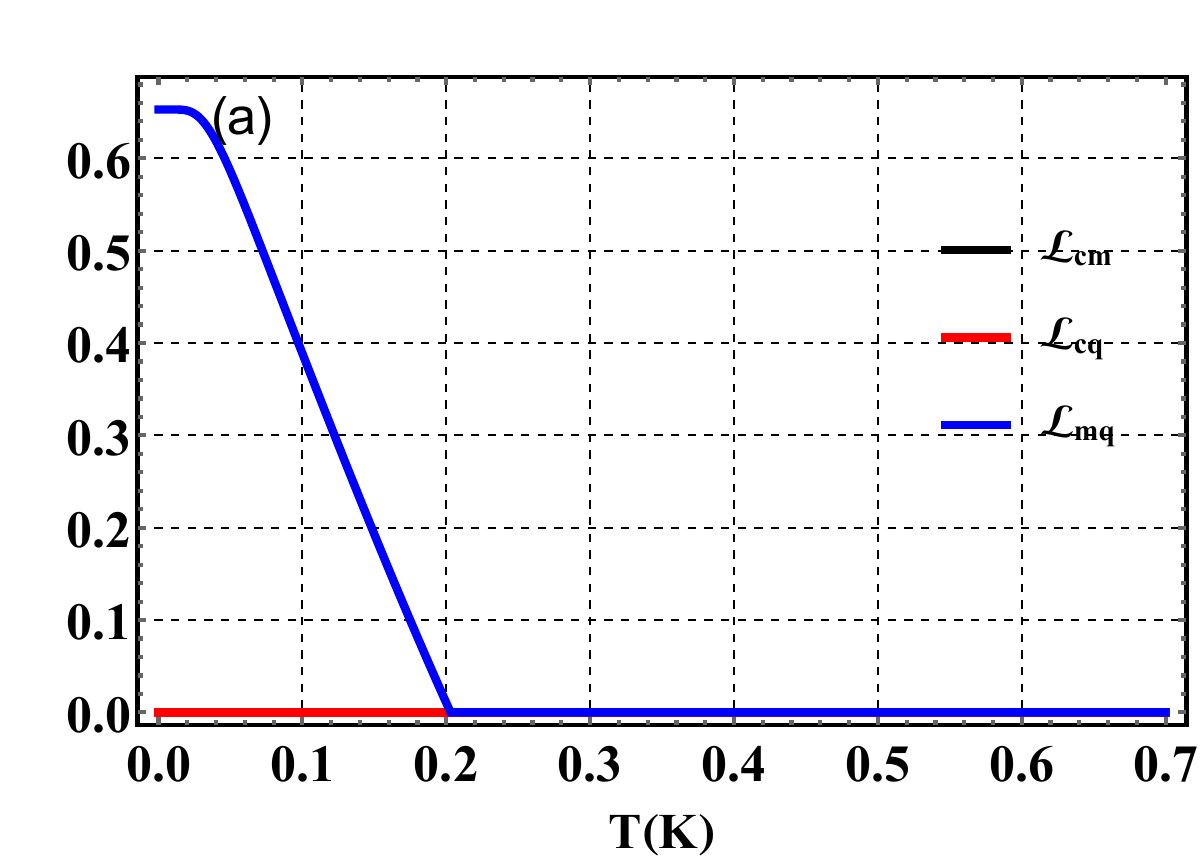} 
\includegraphics[scale=0.35]{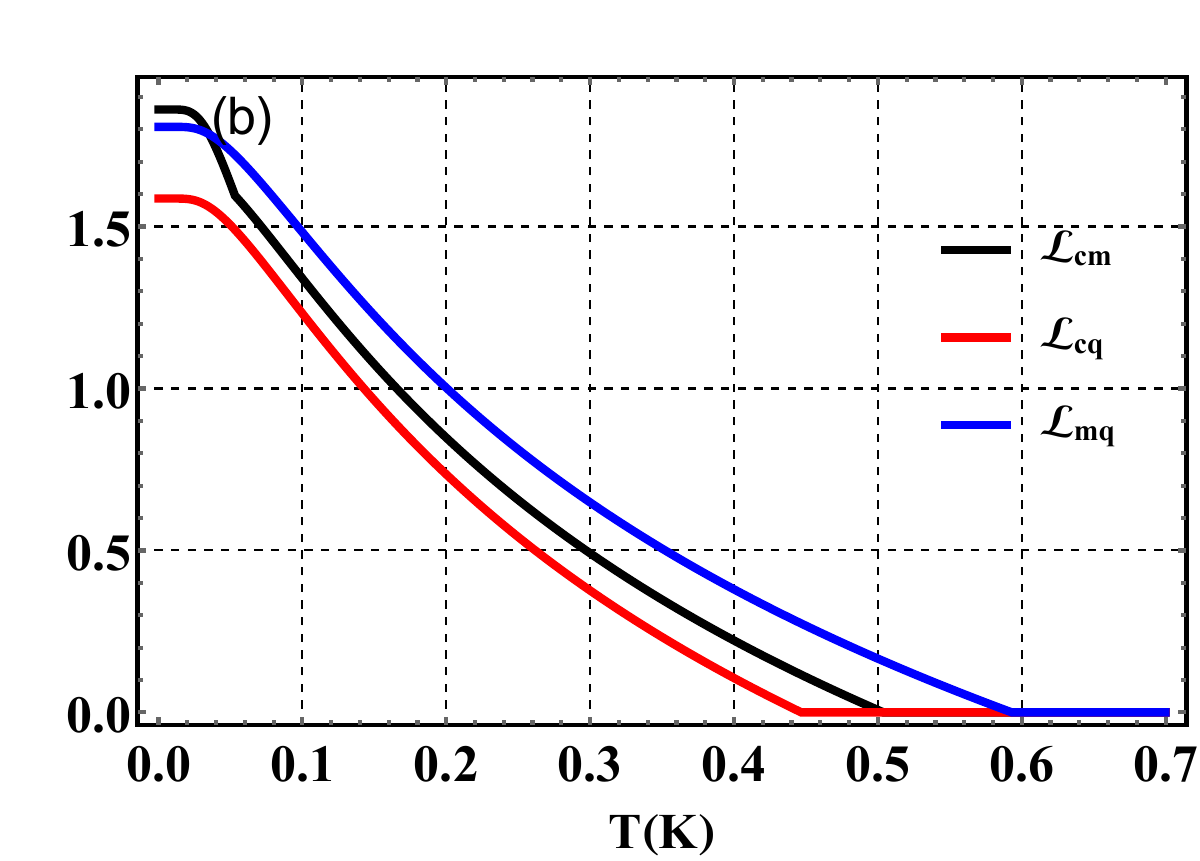} 
\end{tabular}
\caption{Plot of the bipartite entanglement (logarithmic negativity, $\mathcal{L}$) for three mode pairs---cavity-magnon ($\mathcal{L}_{cm}$), cavity-qubit ($\mathcal{L}_{cq}$), and magnon-qubit ($\mathcal{L}_{mq}$)---as a function of temperature $T$. The subfigures correspond to two different reflectivity values: (a) $\epsilon=0$ and (b) $\epsilon=0.86$. The other parameters are set to $\Gamma/2\pi=0.2\times10^6$ Hz and $g_q/\tilde{g}_m=2$.}
\label{fig3}
\end{figure} 

Figure \ref{fig3} indicates the variation of steady-state entanglement with temperature $T$. The plots show the logarithmic negativity for the cavity-magnon ($\mathcal{L}_{cm}$), cavity-qubit ($\mathcal{L}_{cq}$), and magnon-qubit ($\mathcal{L}_{mq}$) mode pairs. The results are shown for the absence of coherent feedback in subfigure (a) and the presence of coherent feedback in subfigure (b). When there is no coherent feedback ($\epsilon=0$), we observe that only the entanglement between the magnon and qubit ($\mathcal{L}_{mq}$) is present, and only at low temperatures. This entanglement vanishes when the temperature $T$ reaches $0.2 \text{ K}$, a result that is consistent with Ref.~\cite{duc1}. The entanglement between the cavity and magnon ($\mathcal{L}_{cm}$) remains zero for all values of $T$, which indicates that these two modes are in a decoherent state. The same is true for the entanglement between the cavity and the qubit ($\mathcal{L}_{cq}$). For $\epsilon=0.86$, we observe an enhancement of the logarithmic negativity for the magnon-qubit modes, with the entanglement persisting up to $T>0.6\text{ K}$. Furthermore, we find the occurrence of entanglement between both the cavity-magnon and cavity-qubit modes. This result confirms that coherent feedback is effective at improving entanglement between the cavity and other system elements through the re-injection of photons. \\

\begin{figure}[ht!]
\centering
\begin{tabular}{ccc}
\includegraphics[scale=0.55]{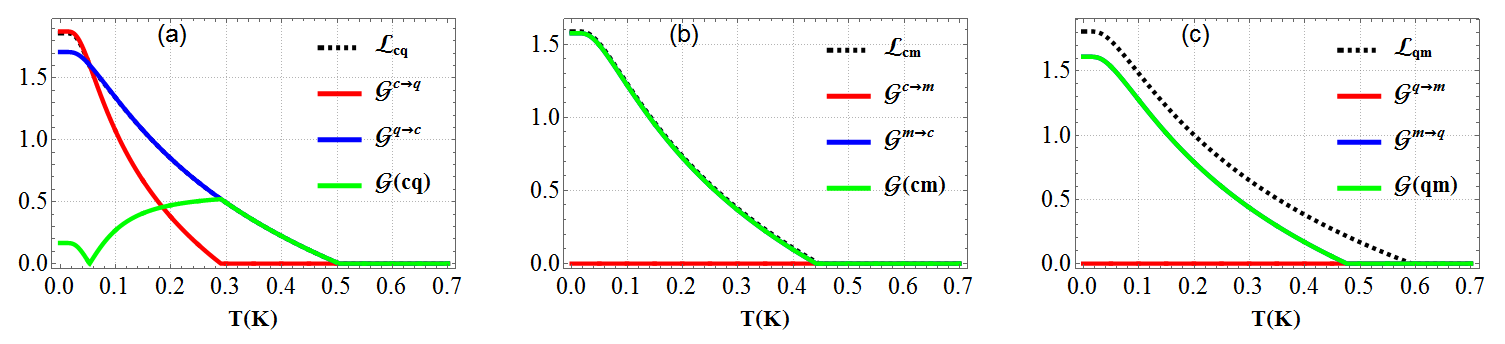} 
\end{tabular}
\caption{Plot of bipartite entanglement, Gaussian quantum steering, and asymmetric quantum steering as a function of temperature $T$. The subfigures display the results for: (a) the cavity-qubit pair ($\mathcal{L}_{cq}$, $\mathcal{G}^{c\to q}$, $\mathcal{G}^{q\to c}$, and $\mathcal{G}(cq)$), (b) the cavity-magnon pair ($\mathcal{L}_{cm}$, $\mathcal{G}^{c\to m}$, $\mathcal{G}^{m\to c}$, and $\mathcal{G}(cm)$), and (c) the qubit-magnon pair ($\mathcal{L}_{qm}$, $\mathcal{G}^{q\to m}$, $\mathcal{G}^{m\to q}$, and $\mathcal{G}(qm)$). The parameters are set to $\Gamma/2\pi=0.2\times10^6$ Hz, $\epsilon=0.86$, $g_q/\tilde{g}_m=2$ and $\theta=\pi$.}
\label{fig2}
\end{figure}

We plot the bipartite entanglement, Gaussian quantum steering ($X\to Y, Y\to X$), and asymmetric steering for the following pairs: (a) cavity-qubit, (b)cavity-magnon, and (c) qubit-magnon. The results, shown as a function of temperature $T$ in Fig.~\ref{fig2}, demonstrate that as the temperature increases, decoherence causes both entanglement and steerability to decrease rapidly. One-way quantum steering is observed to be more robust than two-way steering, persisting at higher temperatures. Although a steerable state must be entangled, the converse is not always true. The presence of Gaussian two-way steering is indicated by the conditions $\mathcal{G}^{X\rightarrow Y}=\mathcal{G}^{Y\rightarrow X}>0$ and $\mathcal{L}_{N}>0$, which confirm that the two subsystems are entangled and mutually steerable. No-way steering emerges when the temperature exceeds $0.5 \text{ K}$ for cavity-qubit modes, $0.45 \text{ K}$ for cavity-magnon and $0.60 \text{ K}$ qubit-magnon modes. The entanglement of a Gaussian state always imposes a limit on its steerability, as discussed in \cite{amazioug}. Furthermore, the asymmetric steering $\mathcal{G}(XY)$ is always bounded by $\ln(2)$, reaching its maximum in cases of one-way steering (i.e., when either $\mathcal{G}^{X\rightarrow Y}>0$ and $\mathcal{G}^{Y\rightarrow X}=0$, or vice versa). As the steerability in either direction increases, the asymmetric steering $\mathcal{G}(XY)$ decreases, which is a result demonstrated in \cite{IKogias2015}.\\

Figure~\ref{fig2}(a) shows that the logarithmic negativity of the cavity-qubit modes ($\mathcal{L}_{cq}$) and the Gaussian steering ($\mathcal{G}^{c\to q}$) exhibit a similar decrease with increasing temperature $T$. Both values become zero when $T$ exceeds $0.5 \text{ K}$. One-way steering emerges when $T$ is greater than $0.3 \text{ K}$, as indicated by the conditions $\mathcal{G}^{c\to q}=0$ and $\mathcal{G}^{q\to c}>0$. Furthermore, the steerability $\mathcal{G}^{c\to q}$ is initially greater than $\mathcal{G}^{q\to c}$ at $T=0 \text{ K}$. However, as the temperature $T$ increases, this relationship is inverted, and we find that $\mathcal{G}^{q\to c}>\mathcal{G}^{c\to q}$. As shown in Figs.~\ref{fig2}(b,c), two-way steering is absent between the cavity-magnon and qubit-magnon modes, since $\mathcal{G}^{c\to m}=0$ and $\mathcal{G}^{q\to m}=0$ for all temperatures. Figure~\ref{fig2}(b) illustrates that the entanglement ($\mathcal{L}_{cm}$) and steering ($\mathcal{G}^{m\to c}$) of the cavity-magnon pair decrease similarly, both vanishing for $T>0.44\text{ K}$. In contrast, Figure~\ref{fig2}(c) demonstrates that the logarithmic negativity ($\mathcal{L}_{qm}$) is greater than the steerability ($\mathcal{G}^{m\to q}$). The entanglement persists up to $T>0.6\text{ K}$, whereas the steerability vanishes earlier, for $T>0.49\text{ K}$. 

\begin{figure}[ht!]
\centering
\begin{tabular}{ccc}
\includegraphics[scale=0.50]{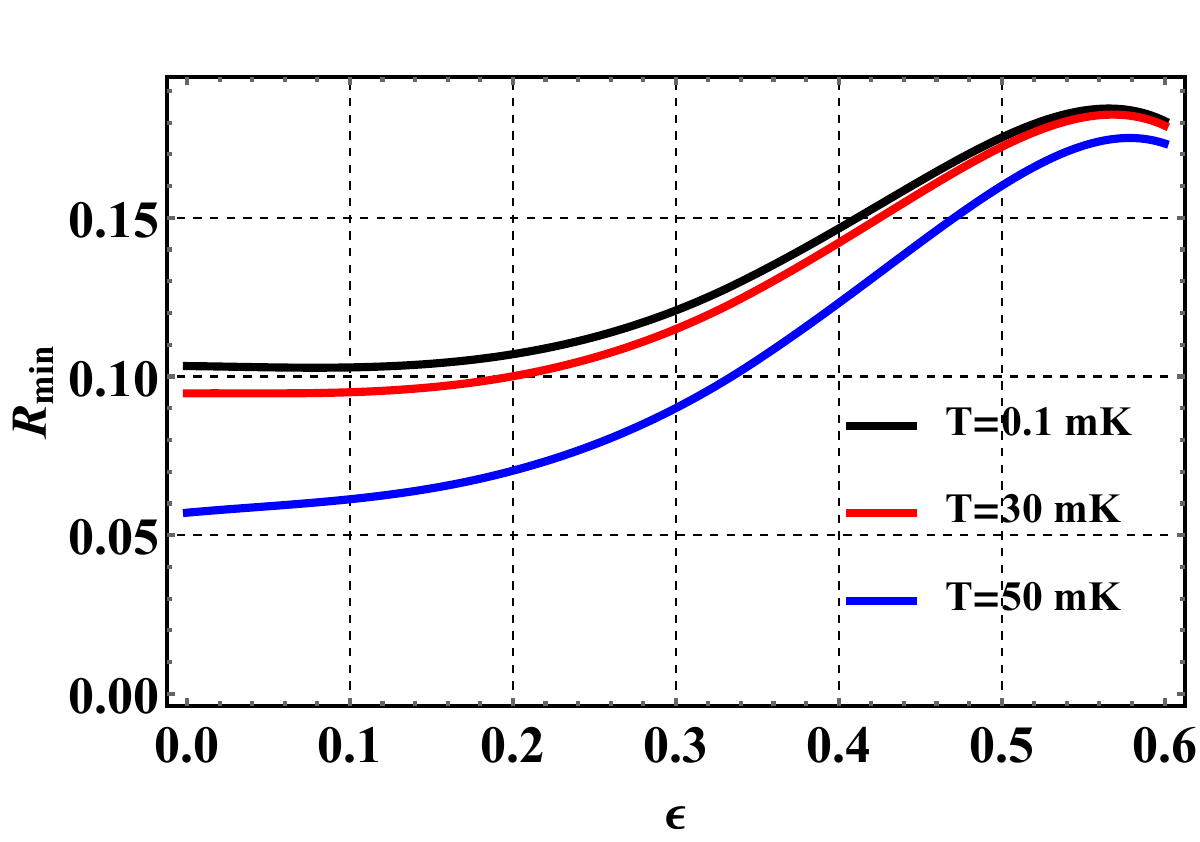} 
\end{tabular}
\caption{Plot tripartite entanglement $\mathbb{R}_{\text{min}}$ as function of the reflictivity parameter $\epsilon$, for different value of temperature $T$, with the parametrers $\Gamma/2\pi=0.2\times 10^{6}$Hz, $g_q/g_0=2$ and $\theta=\pi$.}
\label{fig5}
\end{figure}
Figure \ref{fig5} shows the variation of tripartite entanglement, measured by the minimum residual contangle ($\mathbb{R}_{\text{min}}$), as a function of the reflectivity parameter $\epsilon$ for various temperatures $T$. The figure demonstrates the tripartite entanglement among the three modes (magnon, photon, and qubit). For low reflectivity values ($\epsilon < 0.3$), we observe a partial enhancement of $\mathbb{R}_{\text{min}}$, which decreases as $T$ increases. However, when $\epsilon > 0.3$, the contangle $\mathbb{R}_{\text{min}}$ increases exponentially. We also notice the convergence of the two curves for $T = 0.1$ mK and $T = 30$ mK. This demonstrates that increasing the reflectivity $\epsilon$ enhances the variation of the contangle with temperature.
\begin{figure}[ht!]
\centering
\begin{tabular}{ccc}
\includegraphics[scale=0.50]{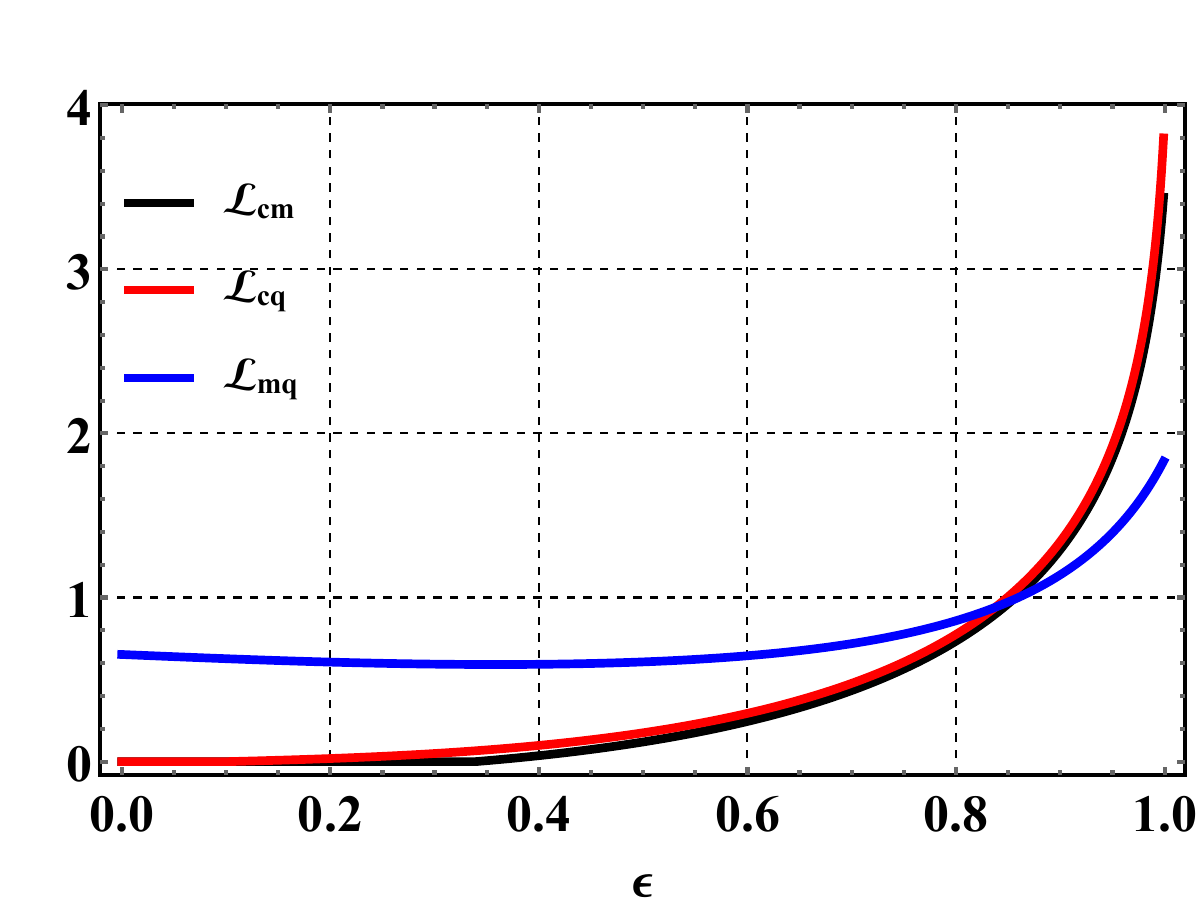} 
\end{tabular}
\caption{Plot of bipartite entanglement (logarithmic negativity, $\mathcal{L}$) for three mode pairs---cavity-magnon ($\mathcal{L}_{cm}$), cavity-qubit ($\mathcal{L}_{cq}$), and magnon-qubit ($\mathcal{L}_{mq}$)---as a function of the reflection coefficient $\epsilon$. The parameters used are $\Gamma/2\pi=0.2\times10^6$ Hz and $\theta=\pi$.}
\label{fig0}
\end{figure}

Figure \ref{fig0} displays how the logarithmic negativities ($\mathcal{L}_{\mathsf{c}\mathsf{m}}$, $\mathcal{L}_{\mathsf{c}\mathtt{q}}$, and $\mathcal{L}_{\mathsf{m}\mathtt{q}}$) change with the reflectivity parameter $\epsilon$. This figure effectively demonstrates the enhancing effect of coherent feedback on entanglement. For reflectivity values below $\epsilon=0.4$, all three entanglement measures remain relatively constant. However, as the parameter exceeds this value ($\epsilon>0.4$), both the cavity-magnon ($\mathcal{L}_{\mathsf{c}\mathsf{m}}$) and cavity-qubit ($\mathcal{L}_{\mathsf{c}\mathtt{q}}$) entanglements show an exponential increase. Concurrently, the magnon-qubit entanglement ($\mathcal{L}_{\mathsf{m}\mathtt{q}}$) also exhibits a partial enhancement. The results clearly show that the feedback loop directly contributes to improving the entanglement between the cavity mode and the other modes in the system. This enhancement is explained by the re-injection of photons into the cavity, which strengthens the quantum correlations. 

\begin{figure}[ht!]
\centering
\begin{tabular}{ccc}
\includegraphics[scale=0.25]{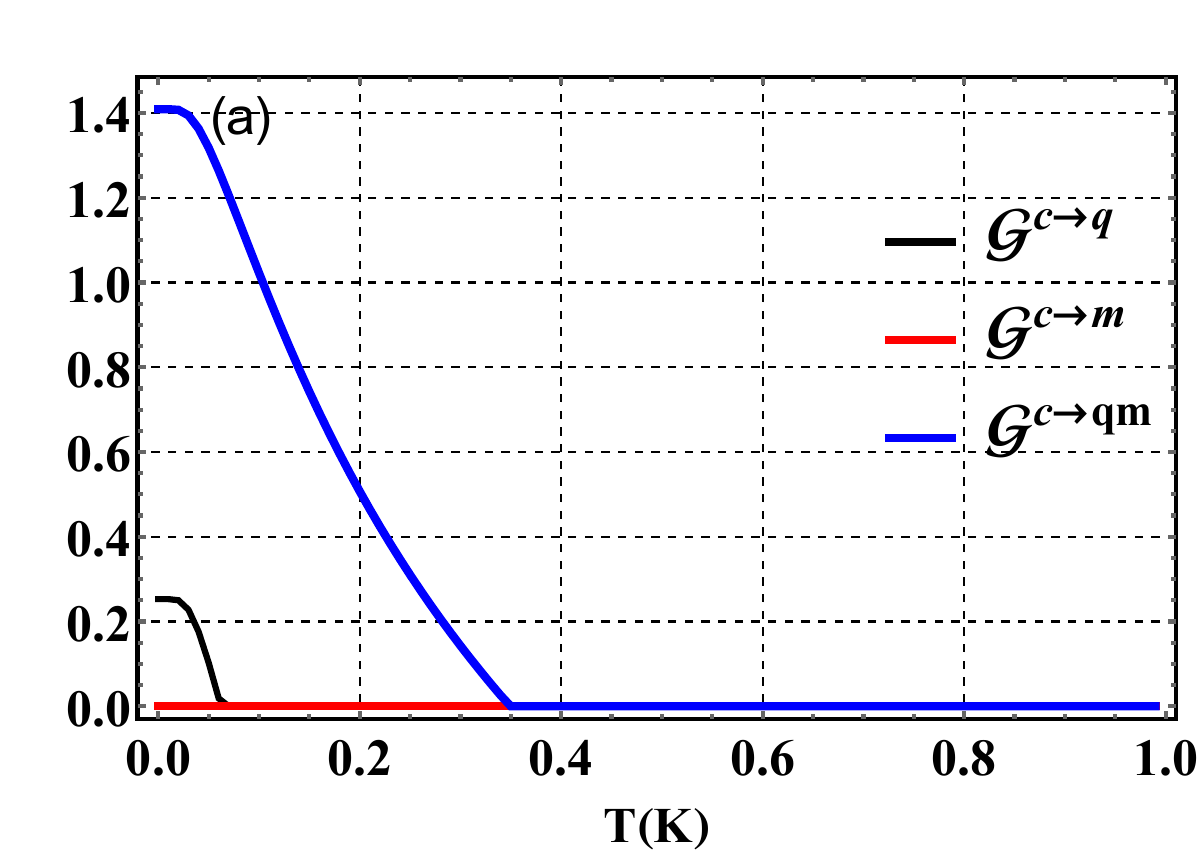}
\includegraphics[scale=0.25]{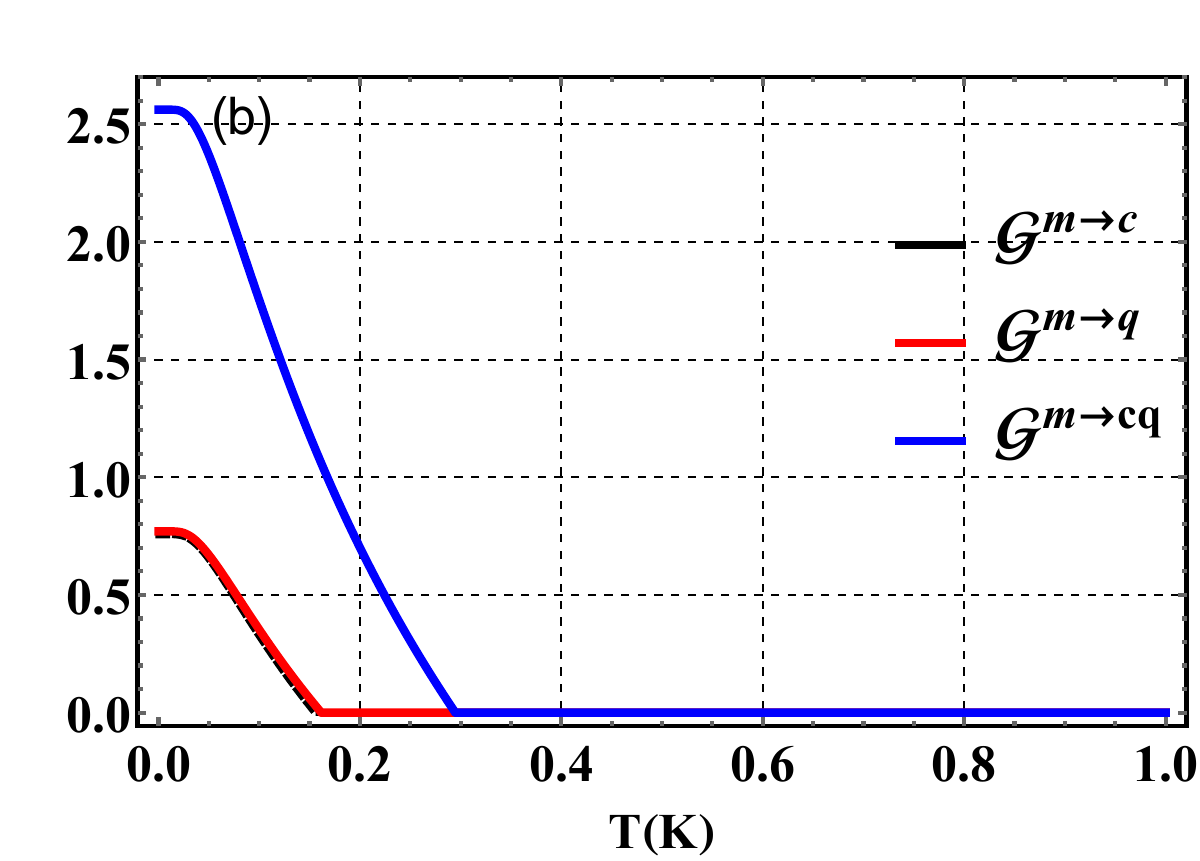}
\includegraphics[scale=0.25]{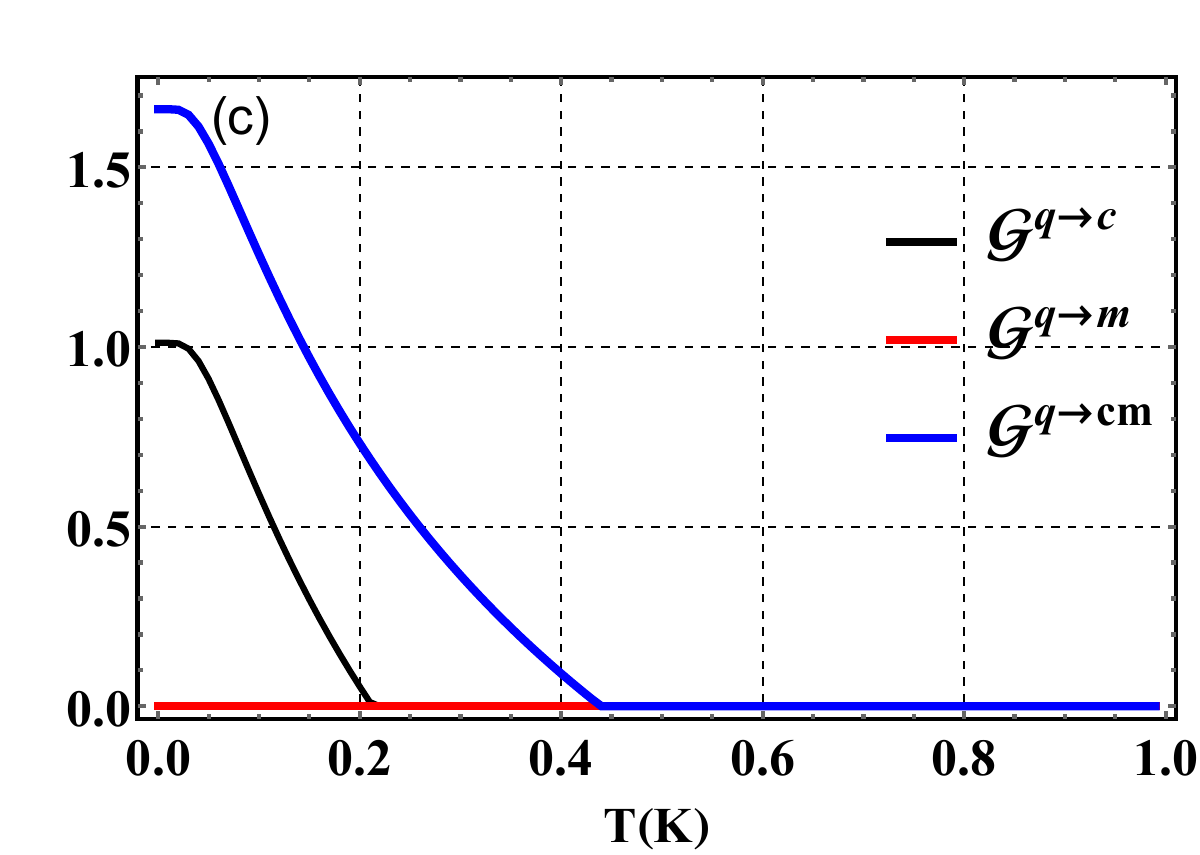} 
\end{tabular}
\caption{Plot of the Gaussian quantum steering as a function of temperature $T$ with a reflection coefficient of $\epsilon=0.90$ and $g_{q}=1.5g_{0}$: (a) steering from the cavity to the qubit ($\mathcal{G}^{c\to q}$), magnon ($\mathcal{G}^{c\to m}$), and qubit-magnon system ($\mathcal{G}^{c\to qm}$); (b) steering from the magnon to the cavity ($\mathcal{G}^{m\to c}$), qubit ($\mathcal{G}^{m\to q}$), and cavity-qubit system ($\mathcal{G}^{m\to cq}$); and (c) steering from the qubit to the cavity ($\mathcal{G}^{q\to c}$), magnon ($\mathcal{G}^{q\to m}$), and cavity-magnon system ($\mathcal{G}^{q\to cm}$).}
\label{fig10}
\end{figure}

Figure \ref{fig10} displays the variation of Gaussian steering as a function of temperature $T$, with a coherent feedback parameter $\epsilon=0.90$. The subfigures show: (a) steering from the cavity to the qubit, magnon, and qubit-magnon system ($\mathcal{G}^{c\to q}$, $\mathcal{G}^{c\to m}$, and $\mathcal{G}^{c\to qm}$); (b) steering from the magnon to the cavity, qubit, and cavity-qubit system ($\mathcal{G}^{m\to c}$, $\mathcal{G}^{m\to q}$, and $\mathcal{G}^{m\to cq}$); and (c) steering from the qubit to the cavity, magnon, and cavity-magnon system ($\mathcal{G}^{q\to c}$, $\mathcal{G}^{q\to m}$, and $\mathcal{G}^{q\to cm}$). We observe that increasing temperature leads to a decrease in the Gaussian steering. Furthermore, in subfigure (\ref{fig10}.a), the steering monogamy relation $\mathcal{G}^{c\to qm}\geq \mathcal{G}^{c\to q}+\mathcal{G}^{c\to m}$ governs the steering among the three modes. The Gaussian steering $\mathcal{G}^{c\rightarrow q}$ is non-zero only at zero temperature, decreasing as $T$ increases and vanishing when $T$ exceeds $0.05 \text{ K}$. Additionally, one-way steering is present for all temperatures, with the condition $\mathcal{G}^{c \rightarrow m}=0$ and $\mathcal{G}^{m \rightarrow c}>0$ holding true, as illustrated in Fig.~\ref{fig10}(b). In Fig.~\ref{fig10}(b), we observe that the steerabilities $\mathcal{G}^{m\to c}$ and $\mathcal{G}^{m\to q}$ decrease in a similar manner and vanish when the temperature exceeds $0.15 \text{ K}$. The Gaussian steering $\mathcal{G}^{m\to cq}$ also decreases with increasing $T$ and disappears when $T>0.3\text{ K}$. Throughout this process, the steering monogamy relation is always verified.

In Fig.~\ref{fig10}(c), we remark that the steerability between the qubit and the cavity modes ($\mathcal{G}^{q\to c}$) is greater than zero, while it is zero between the qubit and the magnon modes ($\mathcal{G}^{q\to m}=0$). This demonstrates one-way steering between the qubit and the magnon modes. This figure also highlights how coherent feedback improves steerability between the cavity and the other elements of the system. We see that the steering $\mathcal{G}^{q\to cm}$ decreases as the temperature $T$ increases. It is worth noting that all steering functions vanish at high temperatures due to decoherence. It can be seen that the three curves of steerability satisfy the steering monogamy inequality, which is correctly stated as $\mathcal{G}^{\alpha\to \beta\gamma}\geq\mathcal{G}^{\alpha\to\beta}+\mathcal{G}^{\alpha\to\gamma}$ for ($\alpha,\beta,\gamma=c,q,m$). This is illustrated in Figs. (\ref{fig10})(a), (b), and (c). The inequality holds simultaneously with the presence of one-way steering (in subfigures (a) and (c)) and two-way steerability (in subfigure (b)). As the temperature increases beyond $T>0.14\text{ K}$, all steering measures vanish, resulting in $ \mathcal{G}^{\alpha\to \beta\gamma}=\mathcal{G}^{\alpha\to\beta}=\mathcal{G}^{\beta \to \gamma}=0$. This result demonstrates that high temperature destroys quantum steering, thereby trivializing the monogamy inequality.

\begin{figure}[ht!]
\centering
\begin{tabular}{ccc}
\includegraphics[scale=0.25]{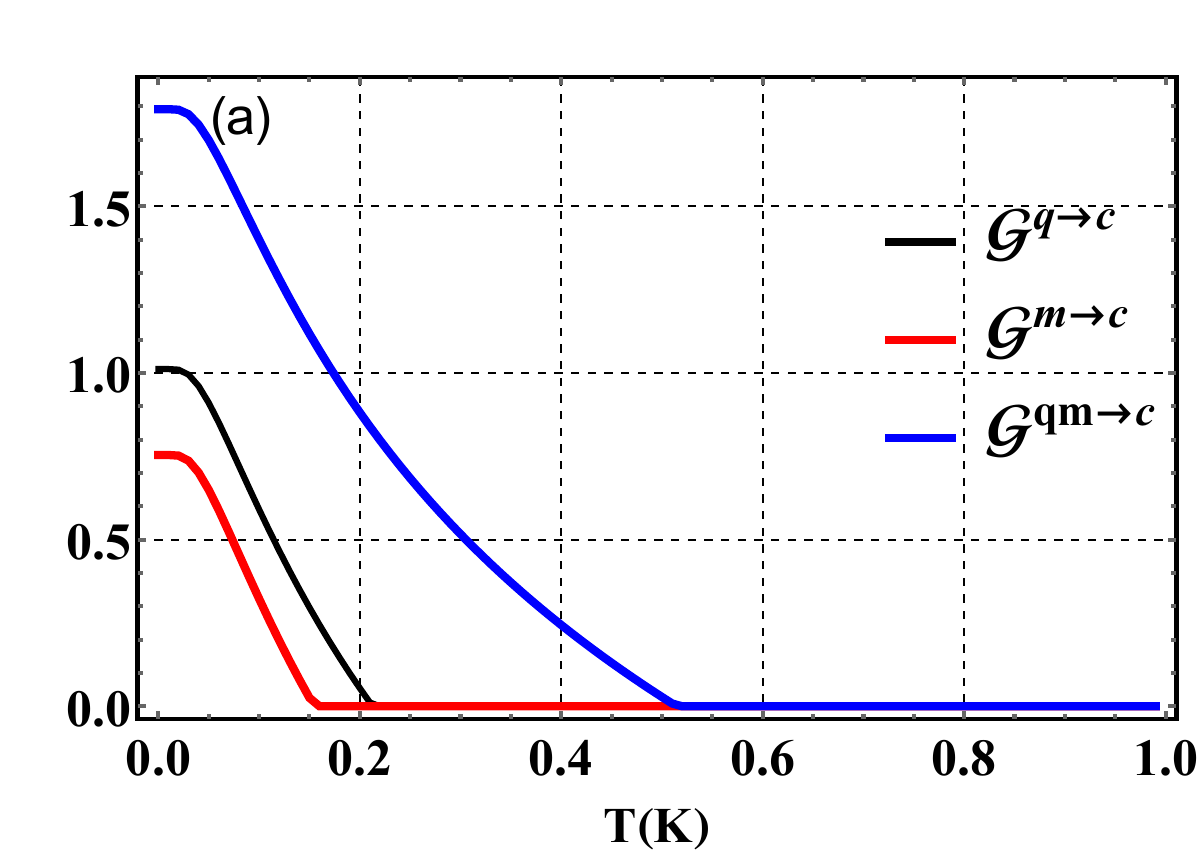}
\includegraphics[scale=0.25]{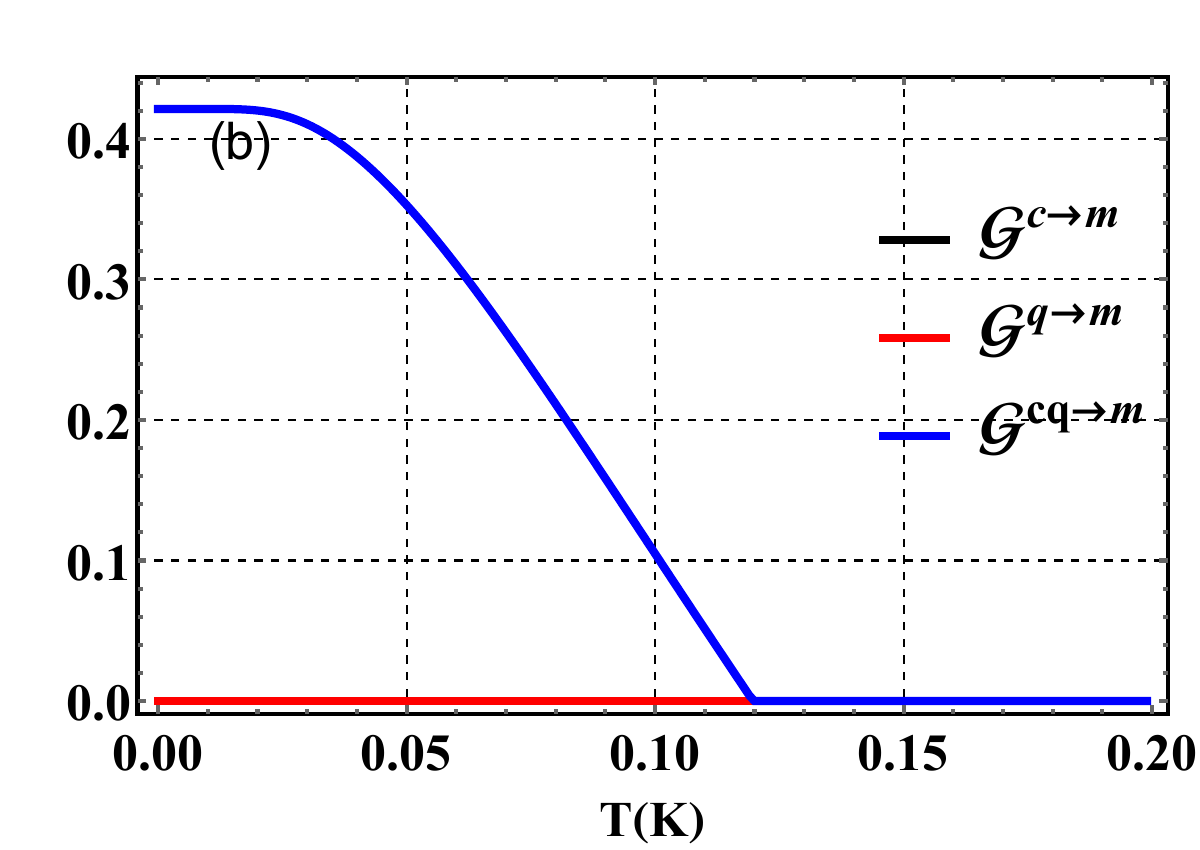} 
\includegraphics[scale=0.25]{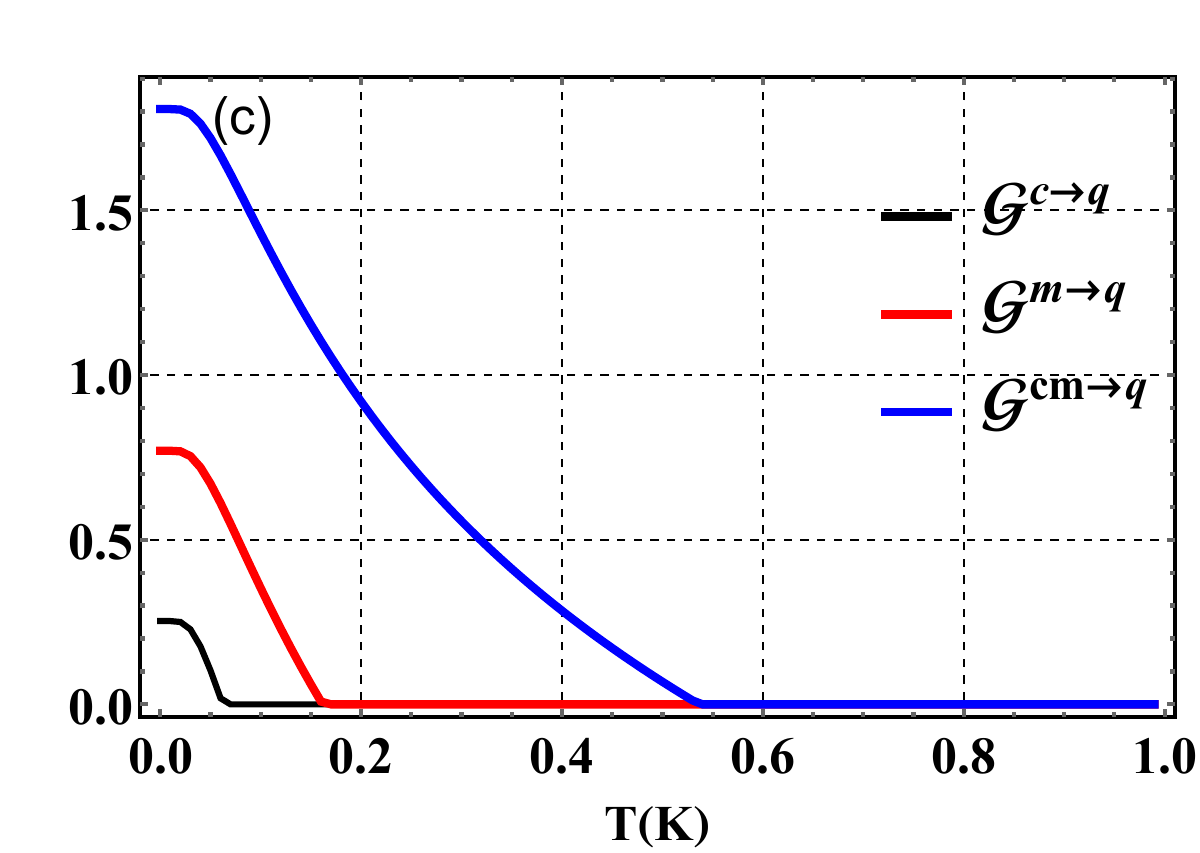}
\end{tabular}
\caption{Plot the steerability as a function of temperature $T$ for a reflection coefficient of $\epsilon=0.90$ and the coupling strength $g_{q}=1.5g_{0}$. The subfigures show steering to a particular mode from other modes or mode pairs: (a) steering to the cavity from the qubit, magnon, and qubit-magnon system ($\mathcal{G}^{q\to c}$, $\mathcal{G}^{m\to c}$, and $\mathcal{G}^{qm\to c}$); (b) steering to the magnon from the cavity, qubit, and cavity-qubit system ($\mathcal{G}^{c\rightarrow m}$, $\mathcal{G}^{q\rightarrow m}$, and $\mathcal{G}^{cq\rightarrow m}$); and (c) steering to the qubit from the cavity, magnon, and cavity-magnon system ($\mathcal{G}^{c\rightarrow q}$, $\mathcal{G}^{m\rightarrow q}$, and $\mathcal{G}^{cm\rightarrow q}$).}
\label{fig11}
\end{figure}
We plot in Fig. \ref{fig11} the Gaussian quantum steering as a function of temperature $T$ in the presence of coherent feedback ($\epsilon=0.90$). The subfigures show steering to a particular mode from other modes or mode pairs: (a) steering to the cavity from the qubit, magnon, and qubit-magnon system ($\mathcal{G}^{q\to c}$, $\mathcal{G}^{m\to c}$, and $\mathcal{G}^{qm\to c}$); (b) steering to the magnon from the cavity, qubit, and cavity-qubit system ($\mathcal{G}^{c\rightarrow m}$, $\mathcal{G}^{q\rightarrow m}$, and $\mathcal{G}^{cq\rightarrow m}$); and (c) steering to the qubit from the cavity, magnon, and cavity-magnon system ($\mathcal{G}^{c\rightarrow q}$, $\mathcal{G}^{m\rightarrow q}$, and $\mathcal{G}^{cm\rightarrow q}$). In Fig.~\ref{fig11}(a), we observe that the steerability $\mathcal{G}^{qm\to c}$ decreases quickly as the temperature $T$ increases. Conversely, the Gaussian steering measures $\mathcal{G}^{q\to c}$ and $\mathcal{G}^{m\to c}$ remain zero for all values of $T$. In Fig.~\ref{fig11}(b), the steerability between the cavity and qubit modes is at its maximum when $T=0 \text{ K}$ and decreases as the temperature increases. We also observe that the Gaussian steering $\mathcal{G}^{c\to m}$ is zero for all values of $T$. This demonstrates the presence of one-way steering between the magnon and cavity modes. In Fig.~\ref{fig11}(c), the Gaussian state obeys the CKW monogamy relation for steering: $\mathcal{G}^{cm\to q}\geq \mathcal{G}^{c\to q}+\mathcal{G}^{m\to q}$. By comparing subfigures, we observe two-way steering between the qubit and cavity modes, as indicated by $\mathcal{G}^{c\to q}>0$ (in subfigure (c)) and $\mathcal{G}^{q\to c}>0$ (in subfigure (a)). We also note the presence of one-way steering between the magnon and qubit modes, since $\mathcal{G}^{m\to q}>0$ while $\mathcal{G}^{q\to m}=0$. This demonstrates that coherent feedback enhances the steerability between the cavity and the other elements of the system. Finally, we show that the steerabilities $\mathcal{G}^{qm\to c}$ and $\mathcal{G}^{cm\to q}$ follow a similar trend, decreasing as the temperature increases and vanishing when $T>0.5\text{ K}$.

\section{Conclusion}   \label{sec5}

In summary, We studied the monogamy of quantum steering and genuine tripartite entanglement in a hybrid qubit-cavity optomagnonic system with a coherent feedback loop. We quantified steady-state entanglement using logarithmic negativity and steerability with Gaussian quantum steering. The genuine tripartite entangled state was measured using the minimum residual contangle. We found that while thermal noise makes both bipartite and tripartite quantum correlations fragile, coherent feedback effectively mitigates these effects. Our verification of the CKW-type monogamy inequalities confirms that steering is monogamous among the three tripartite modes, as shown by adjusting the reflective parameter versus temperature. Ultimately, our results demonstrate that a coherent feedback loop can enhance entanglement and one-way steering even under thermal effects.


\begin{thebibliography}{99}
\bibitem{intro1}
Julsgaard, B., Kozhekin, A., \& Polzik, E. S. Experimental long-lived entanglement of two macroscopic objects. Nature, \textbf{413}(6854), 400-403 (2001).
\bibitem{intro2}
Bowles, J. et al.  Genuinely multipartite entangled quantum states with fully local hidden variable models and hidden multipartite nonlocality. Phys. Rev. Lett. \textbf{116}(13), 130401 (2016).
\bibitem{intro3}
Wang, Z. et al. A flying Schr\"odinger’s cat in multipartite entangled states. Sci. adv. \textbf{8}(10), eabn1778 (2022).
\bibitem{intro4}
Zeng, L. et al. Deterministic distribution of orbital angular momentum multiplexed continuous-variable entanglement and quantum steering. Phot. Res. \textbf{10}(3), 777-785 (2022).
\bibitem{intro5}
Armstrong, S. et al. Multipartite Einstein–Podolsky–Rosen steering and genuine tripartite entanglement with optical networks. N. Phys. \textbf{11}(2), 167-172 (2015).
\bibitem{intro6}
Uola, R. et al. Quantum steering. Rev.  Mod. Phys. \textbf{92}(1), 015001 (2020).
\bibitem{intro7}
Walk, N. et al. Experimental demonstration of Gaussian protocols for one-sided device-independent quantum key distribution. Optica, \textbf{3}(6), 634-642 (2016).
\bibitem{intro8}
Einstein, A., Podolsky, B., \& Rosen, N. Can quantum-mechanical description of physical reality be considered complete?. Phys. Rev. \textbf{47}(10), 777 (1935).
\bibitem{intro9}
Quintino, M. T. et al. Inequivalence of entanglement, steering, and Bell nonlocality for general measurements. Phys. Rev. A \textbf{92}(3), 032107 (2015).
\bibitem{intro10}
Cheng, G., Tan, H., \& Chen, A. Dissipation induced asymmetric steering of distant atomic ensembles. Opt. Commun. \textbf{412}, 166-171 (2018).
\bibitem{intro11}
He, Q., \& Ficek, Z. Einstein-Podolsky-Rosen paradox and quantum steering in a three-mode optomechanical system. Phys. Rev. A \textbf{89}(2), 022332 (2014).
\bibitem{intro12}
Tan, H., Zhang, X., \& Li, G.  Steady-state one-way Einstein-Podolsky-Rosen steering in optomechanical interfaces. Phys. Rev. A \textbf{91}(3), 032121 (2015).
\bibitem{intro13}
Zhan, H., Sun, L., \& Tan, H.  Chirality-induced one-way quantum steering between two waveguide-mediated ferrimagnetic microspheres. Phys. Rev. B \textbf{106}(10), 104432 (2022).
\bibitem{intro14}
Ding, M. S. et al. Enhanced entanglement and steering in PT-symmetric cavity magnomechanics. Opt. Comm. \textbf{490}, 126903 (2021).
\bibitem{intro15}
Jebaratnam, C. et al. Tripartite-entanglement detection through tripartite quantum steering in one-sided and two-sided device-independent scenarios. Phys. Rev. A \textbf{98}(2), 022101 (2018).
\bibitem{intro16}
Zeng, Q. et al. Experimental high-dimensional Einstein-Podolsky-Rosen steering. Phys. Rev. Lett. \textbf{120}(3), 030401 (2018).
\bibitem{intro17}
He, Q. Y., \& Reid, M. D.  Genuine multipartite einstein-podolsky-rosen steering. Phys. Rev. Lett. \textbf{111}(25), 250403 (2013).
\bibitem{intro18}
Vitali, D. et al.  Optomechanical entanglement between a movable mirror and a cavity field. Phys. Rev. Lett. \textbf{98}(3), 030405 (2007).
\bibitem{intro19}
Paternostro, M., Son, W., \& Kim, M. S. Complete conditions for entanglement transfer. Phys. Rev. Lett. \textbf{92}(19), 197901 (2004).
\bibitem{intro20}
Zhang, X.et al.  Cavity magnomechanics sci. Adv. \textbf{2}, e1501286 (2016).
\bibitem{intro21}
Zheng, S. et al. Tutorial: nonlinear magnonics. J.  App. Phys. \textbf{134}15 (2023).
\bibitem{intro22}
Zhang, X. et al. Magnon dark modes and gradient memory. Nat. Comm. \textbf{6}(1), 8914   (2015) .
\bibitem{intro23}
Wang, Y. P. et al. Bistability of cavity magnon polaritons. Phys. Rev. Lett. \textbf{120}(5), 057202 (2018).
\bibitem{thermo24} Amghar, M. B., \& Amazioug, M. Detection of entanglement by harnessing extracted work in magnomechanics. Optik, 311, 171940 (2024).
\bibitem{PLA24} Chabar, N., Amghar, M., \& Amazioug, M. Enhanced Gaussian interferometric power, entanglement and Gaussian quantum steering in magnonics system with squeezed light. Physics Letters A, 519, 129712 (2024).
\bibitem{Fan24} Fan, Z. Y., Zhu, H. B., Li, H. T., \& Li, J. Magnon squeezing via reservoir-engineered optomagnomechanics. APL Photonics, 9(10) (2024).
\bibitem{AdP24} Amazioug, M., Dutykh, D., Teklu, B., \& Asjad, M. Achieving strong magnon blockade through magnon squeezing in a cavity magnetomechanical system. Annalen der Physik, 536(4), 2300357 (2024).
\bibitem{intro24}
Yuan, H. Y. et al. Steady bell state generation via magnon-photon coupling. Phys. Rev. Lett. \textbf{124}(5), 053602 (2020).
\bibitem{Li2018}
Li, J., Zhu, S. Y., \& Agarwal, G. S. Magnon-photon-phonon entanglement in cavity magnomechanics. Phys. Rev. Lett. \textbf{121}(20), 203601 (2018).
\bibitem{intro25}
Sun, F. X. et al. Remote generation of magnon Schrödinger cat state via magnon-photon entanglement. Phys. Rev. Lett. \textbf{127}(8), 087203 (2021).
\bibitem{intro26}
Tabuchi, Y. et al. Coherent coupling between a ferromagnetic magnon and a superconducting qubit. Science, \textbf{349}(6246), 405-408 (2015).
\bibitem{intro30}
Zhang, K. K., Zhu, Z., Shui, T., \& Yang, W. X. Generation of quantum entanglement and Einstein–Podolsky–Rosen steering in a hybrid qubit-cavity optomagnonic system. Chin. J. Phys. \textbf{92}, 284-297 (2024).
\bibitem{intro27}
Mukhopadhyay, D., Nair, J. M., \& Agarwal, G. S. Quantum amplification of spin currents in cavity magnonics by a parametric drive induced long-lived mode. Phys. Rev. B \textbf{106}(18), 184426 (2022).
\bibitem{intro28}
Qi, S. F., \& Jing, J. Magnon-assisted photon-phonon conversion in the presence of structured environments. Phys. Rev. A \textbf{103}(4), 043704 (2021).
\bibitem{intro29}
Haigh, J. A., Nunnenkamp, A., Ramsay, A. J., \& Ferguson, A. J. Triple-resonant Brillouin light scattering in magneto-optical cavities. Phys. Rev. Lett. \textbf{117}(13), 133602 (2016).
\bibitem{Harwood2021}
Harwood, A., Brunelli, M., \& Serafini, A. Cavity optomechanics assisted by optical coherent feedback. Phys. Rev. A, \textbf{103}(2), 023509 (2021).
\bibitem{model1}
Lachance-Quirion, D. et al. Entanglement-based single-shot detection of a single magnon with a superconducting qubit. Science, \textbf{367}(6476), 425-428 (2020).
\bibitem{Ning2021}
Ning, C. X., \& Yin, M.  Entangling magnon and superconducting qubit by using a two-mode squeezed-vacuum microwave field. J. Opt. Soc. Am. B: Opt. Phys. \textbf{38}(10), 3020-3026 (2021).
\bibitem{model2}
Xu, W. L. et al. Magnon-induced optical high-order sideband generation in hybrid atom-cavity optomagnonical system. Opt. Exp. \textbf{28}(15), 22334-22344 (2020).
\bibitem{model3}
Haigh, J. A., Nunnenkamp, A., Ramsay, A. J., \& Ferguson, A. J. Triple-resonant Brillouin light scattering in magneto-optical cavities. Phys. Rev. Lett. \textbf{117}(13), 133602 (2016).
\bibitem{model4}
Gao, Y. P. et al. Magnons scattering induced photonic chaos in the optomagnonic resonators. Nanophotonics, \textbf{9}(7), 1953-1961 (2020).
\bibitem{model5}
Osada, A. et al. Cavity optomagnonics with spin-orbit coupled photons. Phys. Rev. Lett. \textbf{116}(22), 223601 (2016).
\bibitem{model6}
Sun, F. X. et al. Remote generation of magnon Schrödinger cat state via magnon-photon entanglement. Phys. Rev. Lett. \textbf{127}(8), 087203 (2021).
\bibitem{model7}
Holstein, T., \& Primakoff, H. Field dependence of the intrinsic domain magnetization of a ferromagnet. Phys. Rev. \textbf{58}(12), 1098.\bibitem{model8}
Li, J., Zhu, S. Y., \& Agarwal, G. S. (2019). Squeezed states of magnons and phonons in cavity magnomechanics. Phys. Rev. A \textbf{99}(2), 021801 (1940).
\bibitem{model9}
Zhang, W. et al. Quantum entanglement and one-way steering in a cavity magnomechanical system via a squeezed vacuum field. Optics Express, \textbf{30}(7), 10969-10980 (2022).
\bibitem{model10}
Serga, A. A., Chumak, A. V., \& Hillebrands, B. YIG magnonics. J. Phys. D: Appl. Phys. \textbf{43}(26), 264002  (2010).
\bibitem{model11}
Haigh, J. A. et al. Triple-resonant Brillouin light scattering in magneto-optical cavities. Phys. Rev. Lett. \textbf{117}(13), 133602 (2016).
 \bibitem{model16}
Amazioug, M., Maroufi, B., \& Daoud, M. Using coherent feedback loop for high quantum state transfer in optomechanics. Phys. Lett. A \textbf{384}(27), 126705 (2020).
\bibitem{FBAdP}
Sohail, A., Amazioug, M., Singh, S. K., Chabar, N., Ahmed, R., \& de Oliveira, M. C. Coherent Feedback Control of Indirectly Coupled Mode Multipartite Entanglement in a Cavity Opto‐Magnomechanical System. Annalen der Physik, 2400375.

\bibitem{model12}
Viola Kusminskiy, S., Tang, H. X., \& Marquardt, F. Coupled spin-light dynamics in cavity optomagnonics. Phys. Rev. A \textbf{94}(3), 033821 (2016).
\bibitem{model13}
Sun, F. X. et al. Remote generation of magnon Schr\"odinger cat state via magnon-photon entanglement. Phys. Rev. Lett. \textbf{127}(8), 087203 (2021).
\bibitem{model14}
Gardiner, C. W., \& Collett, M. J. Input and output in damped quantum systems: Quantum stochastic differential equations and the master equation. Phys. Rev. A \textbf{31}(6), 3761 (1985).
\bibitem{amazioug2023enhancement}
Amazioug, M., Teklu, B., \& Asjad, M. Enhancement of magnon–photon–phonon entanglement in a cavity magnomechanics with
coherent feedback loop. Scien. Rep. \textbf{13}(1), 3833  (2023).
\bibitem{Li2017}
Li, J. et al. Enhanced entanglement of two different mechanical resonators via coherent feedback. Phys. Rev. A, \textbf{95}(4), 043819 (2017).
\bibitem{model15}
Vitali, D. et al. Optomechanical entanglement between a movable mirror and a cavity field. Phys. Rev. Lett. \textbf{98}(3), 030405(2007).
\bibitem{model17}
Vidal, G., \& Werner, R. F. Computable measure of entanglement. Phys. Rev. A \textbf{65}(3), 032314 (2002).
 \bibitem{model18}
 Adesso, G., Serafini, A., \& Illuminati, F. Extremal entanglement and mixedness in continuous variable systems. Phys. Rev. \textbf{70}(2), 022318 (2004).
 \bibitem{adesso2007entanglement}
Adesso, G., \& Illuminati, F.  Entanglement in continuous-variable systems: recent advances and current perspectives. J.
Phys. A \textbf{40}(28), 7821 (2007).
\bibitem{adesso2006continuous}
Adesso, G., \& Illuminati, F. Continuous variable tangle, monogamy inequality, and entanglement sharing in Gaussian states of
continuous variable systems. New J. Phys. \textbf{8}(1), 15 (2006).
\bibitem{intro31}
Reid, M. D. Monogamy inequalities for the Einstein-Podolsky-Rosen paradox and quantum steering. Phys. Rev. A—Atomic, Molecular, and Optical Phys. \textbf{88}(6), 062108 (2013).
\bibitem{IKogias2015} Kogias, I., Lee, A. R., Ragy, S., \& Adesso, G. Quantification of Gaussian quantum steering. Phys. Rev. Lett. \textbf{114}(6), 060403 (2015).
\bibitem{model21}
Zhan, Z. W. et al. Coupled three-mode squeezed vacuum: Gaussian steering and remote generation of Wigner negativity. Phys. Rev. A \textbf{108}(1), 012436 (2023).
\bibitem{VCoffman2000}
Coffman, V., Kundu, J., \& Wootters, W. K. Distributed entanglement. Phys. Rev. A \textbf{61}(5), 052306 (2000).
\bibitem{YXiang2017}
Xiang, Y.et al. Multipartite Gaussian steering: Monogamy constraints and quantum cryptography applications. Phys. Rev. A \textbf{95}(1), 010101 (2017).
\bibitem{model22}
Rueda, A. et al. Electro-optic entanglement source for microwave to telecom quantum state transfer. npj Quan. Inf. \textbf{5}(1), 108  (2019).
\bibitem{model23}
Sun, F. X. et al. Remote generation of magnon Schrödinger cat state via magnon-photon entanglement. Phys. Rev. Lett. \textbf{127}(8), 087203  (2021).
\bibitem{model24}
Lachance-Quirion, D. et al. Entanglement-based single-shot detection of a single magnon with a superconducting qubit. Science, \textbf{367}(6476), 425-428 (2020).
\bibitem{duc1}
Zhang, K. K., Zhu, Z., Shui, T., \& Yang, W. X. Generation of quantum entanglement and Einstein–Podolsky–Rosen steering in a hybrid qubit-cavity optomagnonic system. Chin. J. Phys. \textbf{92}, 284-297  (2024).
\bibitem{amazioug}
Amazioug, M., Nassik, M., \& Habiballah, N.  Gaussian quantum discord and EPR steering in optomechanical system. Optik, \textbf{158}, 1186-1193 (2018).


\end{thebibliography}
\end{document}